\documentclass[11pt,epsf,letterpaper]{article}%
\usepackage{setspace}
\usepackage{color}
\usepackage{graphicx}
\usepackage{amsmath}
\usepackage{amsfonts}
\usepackage{verbatim}
\usepackage{amssymb}

\usepackage{url}

\usepackage[title]{appendix}

\providecommand{\U}[1]{\protect\rule{.1in}{.1in}}
\newcommand{\be}{\begin{equation}}
\newcommand{\ee}{\end{equation}}
\newcommand{\ba}{\begin{eqnarray}}
\newcommand{\ea}{\end{eqnarray}}

\newcommand{\drm}{{\rm d}}

\begin{document}

\title{\bf A road map for Feynman's adventures in the land of gravitation}

\author{Marco Di Mauro$^{1}$,
Salvatore Esposito$^{2}$,
Adele Naddeo$^{2}$ \\
$^{1}$\footnotesize{Dipartimento di Matematica, Universit\`a di
Salerno,
Via Giovanni Paolo II, 84084 Fisciano, Italy.}\\
$^{2}$\footnotesize{INFN Sezione di Napoli, Via Cinthia, 80126
Naples, Italy.}} \maketitle

\begin{abstract}

\noindent Richard P. Feynman's work on gravitation, as can be
inferred from several published and unpublished sources, is
reviewed. Feynman was involved with this subject at least from
late 1954 to the late 1960s, giving several pivotal contributions
to it. Even though he published only three papers, much more
material is available, beginning with the records of his many
interventions at the Chapel Hill conference in 1957, which are
here analyzed in detail, and show that he had already considerably
developed his ideas on gravity. In addition he expressed deep
thoughts about fundamental issues in quantum mechanics which were
suggested by the problem of quantum gravity, such as
superpositions of the wave functions of macroscopic objects and
the role of the observer. Feynman also lectured on gravity several
times. Besides the famous lectures given at Caltech in 1962-63, he
extensively discussed this subject in a series of lectures
delivered at the Hughes Aircraft Company in 1966-67, whose focus
was on astronomy and astrophysics. All this material allows to
reconstruct a detailed picture of Feynman's ideas on gravity and
of their evolution until the late sixties. According to him,
gravity, like electromagnetism, has quantum foundations, therefore
general relativity has to be regarded as the classical limit of an
underlying quantum theory; this quantum theory should be
investigated by computing physical processes, as if they were
experimentally accessible. The same attitude is shown with respect
to gravitational waves, as is evident also from an unpublished
letter addressed to Victor F. Weisskopf. In addition, an original
approach to gravity, which closely mimics (and probably was
inspired by) the derivation of the Maxwell equations given by
Feynman in that period, is sketched in the unpublished Hughes
lectures.
\end{abstract}
\begin{flushright}
\emph{Dedicated to the Memory of Erasmo Recami.}
\end{flushright}
\section{Introduction}

Richard P. Feynman's approach to physics was not a sectorial one.
According to his vision, all branches of it are parts of a whole,
reflecting the innermost unity of nature itself\footnote{For
example, in \cite{Feynman:1963uxa}, chap. I-3, he says: ``If our
small minds, for some convenience, divide this glass of wine, this
universe, into parts-physics, biology, geology, astronomy,
psychology, and so on-remember that nature does not know it! So
let us put it all back together, not forgetting ultimately what it
is for'', while he explicitly refers to ``the underlying unity of
nature'' in chap II-12.}. In particular, he was among the first to
be concerned about the relation of gravitation to the rest of
physics\footnote{``The problem of the relation of gravitation to
the rest of physics is one of the outstanding theoretical problems
of our age''(\cite{ChapelHill}, p. 15); ``My subject is the
quantum theory of gravitation. My interest in it is primarily in
the relation of one part of nature to another''
(\cite{Feynman:1963ax}, p. 697).}, in a period (the early 1950s)
where general relativity practitioners still tended to be isolated
from mainstream research, while the so-called \emph{renaissance}
of general relativity would only take place some years
later\footnote{This expression was coined by Clifford M. Will in
his popular book \cite{Will} (see also \cite{Will:1989rr}), to
denote the period going roughly from the mid-late fifties to the
late seventies, in which general relativity gradually switched
from being a subject at the margins of physics (as it was from the
mid-twenties to the mid-fifties, a period which Jean Eisenstaedt
\cite{Eisenstaedt1, Eisenstaedt2,Eisenstaedt3} called the
\emph{low water-mark} of general relativity) to being a mainstream
subject.} \cite{BlumLalliRenn}. Feynman's interest on gravity
dates at least from late 1954 (as recalled in \cite{Gell-Mann,
Letter})\footnote{Since Murray Gell-Mann wrote that Feynman had
``made considerable progress'', it is very likely that he had
already been working on gravity for some years in 1954, probably
starting shortly after taming quantum electrodynamics, in the
early 50s.}. He gave several fundamental contributions in the
following years, until the late sixties, when he apparently lost
interest in the subject. The sticky-bead argument
\cite{ChapelHill}, the Feynman rules for general relativity  and
the associated ghosts \cite{Feynman:1963ax}, the Caltech Lectures
on Gravitation \cite{Feynman:1996kb}, and the
Feynman-Chandrasekhar instability of supermassive stars
\cite{Chandrasekhar:1964zz}, are now part of the common lore about
classical and quantum gravity.

In this paper we retrace the full development of Feynman's ideas
on this subject. We start from the 1957 Chapel Hill conference,
where for the first time Feynman's thoughts on gravity were
publicly expressed and whose written records are widely available
\cite{ChapelHill}. At that conference, which was pivotal in
triggering the renaissance of general relativity, the
gravitational physics community delineated the tracks along which
subsequent work would develop. Besides cosmology, which at the
time was still considered ``a field on its own, at least at
present, not intimately connected with the other aspects of
general relativity'' (\cite{BergmannRMP}, p. 352), the main issues
to be addressed at Chapel Hill were \cite{BergmannRMP}: classical
gravity, quantum gravity, and the classical and quantum theory of
measurement (as a link between the previous two topics). The
records of Feynman's interventions at that conference show that he
had already deeply thought, and performed computations, about all
three topics, focusing on classical gravitational waves, on
arguments in favor of quantum gravity from fundamental quantum
mechanics, and finally on quantum gravity itself. We shall
therefore develop our narrative along these lines, starting with
Feynman's interventions at Chapel Hill and then following the
developments of the subsequent years. Some not well-known
unpublished material is considered as well, namely the
transcriptions of two sets of lectures, which Feynman delivered in
the years 1966-67 \cite{FeynmanHughes1} and 1967-68
\cite{FeynmanHughes2} at the Hughes Aircraft Company, which have
recently been made available on the web\footnote{The story of how
Feynman got involved in teaching at the Hughes Aircraft Company is
briefly told in \cite{Feynmanbio}.} \cite{Hughes}. In particular,
the 1966-67 lectures \cite{FeynmanHughes1}, which were devoted to
astronomy, astrophysics and cosmology, contain a preliminary
discussion of the elements of general relativity. This treatment
displays many similarities with the more famous Lectures on
Gravitation, delivered at Caltech in 1962-63
\cite{Feynman:1996kb}, but also several differences, probably a
consequence of the evolution of the physicist's ideas in the years
between the two courses. In those years, as we reported elsewhere
\cite{DeLuca:2019ija,DiMauro:2020bpd}, Feynman developed a new
derivation of Maxwell's equations, with the aim of finding an
original way of teaching electromagnetism. Probably inspired by
this, in the Hughes lectures he suggested that a similar approach,
with suitable modifications, could be adopted also for gravity.
This suggestion is scattered in several places in the lectures on
astrophysics \cite{FeynmanHughes1}, but also in those given in the
following year (1967-68), whose focus was on electromagnetism
\cite{FeynmanHughes2}. Feynman limited himself to these hints,
without pursuing them further, plausibly because of the
considerably higher analytical complexity of developing full
general relativity along these lines, in comparison with
electromagnetism.

The paper is organized as follows: in Section 2 we introduce the
Chapel Hill conference and Feynman's contributions to it. In
Section 3 we focus on gravitational waves and on the sticky-bead
argument. In Section 4 we analyze Feynman's considerations on the
foundational issues related to the quantization of gravity. In
Section 5 we discuss and put into context his arguments in favor
of a field theoretical viewpoint on classical general relativity
and on its subsequent quantization. In Section 6 we describe his
work on the quantization and renormalization of gravity. In
Section 7 we give an overview of the parts of the Hughes lectures
of 1966-67 \cite{FeynmanHughes1} focusing on relativistic gravity
issues, and we draw a comparison with the treatment given in the
Lectures on Gravitation \cite{Feynman:1996kb}. Finally, we discuss
the approach to gravity sketched both in the 1966-67
\cite{FeynmanHughes1} and in the 1967-68 Hughes lectures
\cite{FeynmanHughes2}. Section 8 is devoted to our conclusions.

\section{The Chapel Hill conference}

The renaissance of general relativity was characterized by the
establishment of a community of researchers, which was also
achieved through the organization of a series of international
conferences entirely devoted to the subject \cite{BlumLalliRenn}.
The first one was the Bern Jubilee conference of 1955 \cite{Bern},
celebrating the fiftieth anniversary of the formulation of special
relativity. The Chapel Hill conference, organized in 1957 by Bryce
S. DeWitt and his wife C\'ecile DeWitt-Morette \cite{ChapelHill},
was the second one (although it is commonly referred to as GR1,
while the Bern conference is called GR0). Unlike the Bern
conference, which involved mostly European physicists of the older
generation, the Chapel Hill conference involved many younger
physicists, in particular many Americans, which would soon become
leaders in the field\footnote{The full participant list is
reported in \cite{ChapelHill}, pp. 39-40.}. In fact, it had a
bigger impact on the field and contributed a great deal to
defining the trends for most of the subsequent research in
classical and quantum gravity, and to recognizing general
relativity as a genuine physical theory. This recognition, whose
need was widely felt among the practitioners, was indeed the main
aim of the conference, as its title (\textit{The Role of
Gravitation in Physics}) clearly shows. An excellent account of
the events that brought to the organization of the conference can
be found in the introduction to the recent republication of the
conference records \cite{ChapelHill}.

\subsection{Feynman's contributions}

During the Chapel Hill conference, Feynman discussed the reality
of gravitational waves, proposing the well-known sticky-bead
argument to simply and intuitively argue that gravitational waves
carry energy, and therefore are not just coordinate artifacts.
Also, he discussed several foundational issues in quantum
mechanics, which were linked with the problem of the quantization
of gravity; in particular, he explicitly characterized Hugh A.
Everett's novel interpretation of quantum mechanics, which was
presented there for the first time\footnote{This topic was briefly
mentioned in a discussion session (\cite{ChapelHill}, p. 270) by
Everett's Ph.D. advisor John A. Wheeler (Everett himself did not
participate to the conference), who also urged DeWitt to include
an account of it (which was an abridged version of Everett's Ph.D.
thesis) in the special issue of the Reviews of Modern Physics
devoted to the conference \cite{Everett} (cf.
\cite{DeWittMorette:2011zz}, p. 94).}, in terms of
``many-worlds''\footnote{This expression was popularized by DeWitt
\cite{DeWittBattelle}-\cite{DeWittGraham}, who in the late sixties
and early seventies undertook the task of making Everett's
interpretation better known among physicists, having realized that
it had been substantially unnoticed after its introduction in 1957
(see \cite{DeWittMorette:2011zz}, pp. 94-97).}, in an attempt to
criticize it (cf. the quotation reported in Section
\ref{Everett}). Lastly, Feynman advocated a quantum field
theoretical approach to general relativity, as opposed to the
geometric one which had been the standard one since 1915. Such an
approach was in fact being pursued and taught by several particle
theorists who got involved in gravity in that period, such as
Sidney Coleman \cite{Kaiser}. Indeed, according to Feynman, the
geometric approach prevented the merging of gravity with the rest
of physics, although he considered it to be beautiful and elegant.
As he later commented:
\begin{quote}
This great argument\footnote{Here Feynman is referring to Albert
Einstein's geometric approach.} is so perfect it has never been
used to solve other physical problems. This is perhaps due to the
geometrical interpretation of gravity which no one can unravel
into non geometrical arguments (\cite{FeynmanHughes1}, p. 30).
\end{quote}
Feynman did not like rigorous mathematical arguments. In his
opinion, in a field like gravity, where experimental input was, at
that time, limited (or even absent for the quantum case),
physicists could rely either on mathematical rigor or on
imagination, that is, on thought experiments (\cite{ChapelHill},
pp. 271-272). Since he did not believe in rigor, he chose the
second way, as is evident in all his work on gravity. In his own
words (\cite{ChapelHill}, p. 272):
\begin{quote}
I think the best viewpoint is to pretend that there are
experiments and calculate. In this field since we are not pushed
by experiments we must be pulled by imagination.
\end{quote}
He would in fact often try to compute measurable (at least in
principle) physical effects, like the Lamb shift of a
gravitationally held atom, the Compton scattering of gravitons, or
the energy dissipated by the Sun-Earth system through radiation of
gravitational waves. His arguments in favor of the quantization of
the gravitational field, and of the reality of gravitational
waves, relied on thought experiments rather than on sophisticated
mathematical analysis. This attitude explains why, unlike most
relativists, he gave up the geometrical approach without much
regret. A second motivation for him to treat gravity in that way
was his belief that quantum mechanics underlies the basic
structure of nature, with the forces we observe at the macroscopic
level emerging from quantum mechanics in the classical limit:
\begin{quote}
I shall call conservative forces, those forces which can be
deduced from quantum mechanics in the classical limit. As you
know, Q.M. is the underpinning of Nature (\cite{FeynmanHughes2},
p.35).
\end{quote}
This is of course true for electromagnetism, as Feynman declared
in several places \cite{DiMauro:2020bpd}, and it is true for
gravity as well. As a matter of fact, even Feynman's
considerations in classical gravity, such as those dealing with
gravitational waves, have their roots in quantum physics, as
classical general relativity was considered the tree diagram
approximation of a quantum theory of gravity.

\section{Do gravitational waves exist?}

The general covariance of general relativity makes it difficult --
in the absence of an invariant characterization -- to distinguish
between real physical phenomena and artifacts due to the choice of
the coordinate system (which would disappear by changing
coordinates). At the time of the Chapel Hill conference, this was
not clear yet for gravitational waves, which until then had been
studied only in special coordinate systems\footnote{For an
excellent account of the fascinating history of gravitational
waves, which adds considerable detail to what follows, see
\cite{Kennefick:2007zz}.}. In 1916 \cite{Einstein:1916cc} and 1918
\cite{Einstein:1918btx}, Einstein first found that his
gravitational field equations, when linearized, admit wave-like
solutions travelling with the speed of light, by adopting a
particular coordinate system. He also showed that at least some of
these solutions carried energy and provided a formula for the
leading-order energy-emission rate, the famous \emph{quadrupole
formula}. However, he encountered some difficulties in the
definition of such energy, due to the fact, already known to him,
that energy is not localizable in general relativity. In 1922,
Arthur S. Eddington noted the weakness of Einstein's derivation:
\begin{quote}
The potentials $g_{\mu\nu}$ pertain not only to the gravitational
influence which has objective reality, but also to the
coordinate-system which we select arbitrarily. We can ``propagate"
coordinate-changes with the \emph{speed of thought}, and these may
be mixed up at will with the more dilatory propagation discussed
above. There does not seem to be any way of distinguishing a
physical and a conventional part in the changes of the
$g_{\mu\nu}$ (\cite{Eddington}, p. 130, emphasis in the
original)\footnote{In fact, Eddington was not questioning (as is
sometimes stated) the existence of gravitational disturbances
travelling with a finite speed, which was required by special
relativity, but rather Einstein's procedure, in which the speed of
light seemed to be put in by hand by the particular coordinate
choice he adopted. In fact, in a subsequent paper
\cite{Eddington:1922ds}, Eddington showed more rigorously that
some of Einstein's wave solutions were not spurious, i.e. were not
just flat space in curvilinear coordinates, and did propagate with
the speed of light. In the same paper, he corrected a minor error
in Einstein's quadrupole formula (a factor of 2), and applied it
to compute the radiation reaction for a rotator.}.
\end{quote}
Further doubts concerned the possibility that gravitational waves
were just artifacts of the linearized approximation, which would
disappear once the full theory was considered. In fact, Einstein
himself \cite{Pais}, while attempting, along with Nathan Rosen, to
find plane wave solutions of the full nonlinear field equations of
gravitation, became convinced that the full theory predicted no
gravitational waves\footnote{See also \cite{EinsteinBorn}, letter
71.}. However, after a mistake was found\footnote{The
singularities that Einstein and Rosen had found were in fact mere
coordinate artifacts \cite{Kennefick:2007zz}.}, Einstein and Rosen
finally discovered a rigorous solution describing waves radiated
off the axis of an infinite cylinder \cite{EinsteinRosen1937}.
This did not settle the issue yet, since the problem of the
physical effects and energy conveying of gravitational waves
remained unsolved. Such problems arose as a consequence of the
lack (at that time) of a general relativistic theory of
measurement, along with the already mentioned difficulties
inherent in the definition of energy in general relativity. In
particular, the wave solutions found by Einstein and Rosen seemed
not to carry energy, as still argued by Rosen himself in 1955
(\cite{Bern}, pp. 171-174), and it was not known whether this
depended on their unbounded nature; also, it was not known whether
full general relativity admitted spherical wave solutions,
describing energy radiated by a localized center.

The dilemma thus persisted that gravitational waves could exist
only as mathematical objects, with no actual physical content, and
all the above problems remained unsolved until the appearance of
Felix A. E. Pirani's work in 1956 \cite{Pirani1,Pirani2} and the
subsequent discussions which took place at Chapel Hill in the
following year \cite{ChapelHill}, where this work was presented.
While investigating the physical meaning of the Riemann tensor in
terms of the geodesic deviation equation, Pirani managed to give
an invariant definition of gravitational radiation, in terms of
spacetime curvature. Gravitational waves would in fact consist in
propagating ripples in spacetime\footnote{More precisely,
gravitational wave-fronts manifest themselves as propagating
discontinuities in the Riemann tensor across null 3-surfaces.}.
Since curvature modifies the proper distance between test
particles, a gravitational wave producing such an effect had to be
real\footnote{This same reasoning was also applied to the
cylindrical waves of Einstein and Rosen by Joseph Weber and
Wheeler in the special issue of Reviews of Modern Physics devoted
to the conference \cite{Weber:1957oib}.}. As we will see, this was
Feynman's starting point when he proposed his famous sticky-bead
argument\footnote{The exact words ``sticky bead'' are usually
attributed to Feynman. However, in the proceedings
\cite{ChapelHill} there is no record of his use of them at Chapel
Hill. We could not retrace the first usage of that expression.}.

An additional unsolved issue, related to the former, came to the
physicists' attention. It concerned the energy-loss rate in a
binary star system due to the emission of gravitational waves, and
whether, in this process, the energy carried by such waves is
proportional to the square of their amplitude. In fact, such a
gravitationally-bound system cannot be described in the linearized
approximation, and it was not known whether Einstein's quadrupole
formula correctly describes the energy loss beyond the linear
approximation. The computation was further complicated by the fact
that the radiation reaction on the emitting stars has to be taken
into account. Indeed, it is very difficult to analytically solve
the Einstein equations in the case of a strongly gravitating
source, in order to compute the amplitude of gravitational waves.
A widely used approximation scheme, suitable to
gravitationally-bound systems, is the \emph{post-Newtonian} one
(for reviews see \cite{postN1,postN2}). It assumes, as small
parameters, the magnitude of the metric deviation from the
background Minkowski metric and the squared ratio
$\left(\frac{u}{c}\right)^{2}$ between the typical velocity of the
system $u$ and the speed of light $c$. For instance, in the case
of a binary system, the typical velocity is the average orbital
velocity while the Newtonian potential measures deviations from
the flat metric at the lowest order. The result is a Newtonian
description at the lowest order, while relativistic effects arise
at higher orders. The radiative effects emerge at the
order\footnote{This is the lowest nontrivial order with an odd
power of $v$, which signals the dissipation of energy due to
radiation.} $\left(\frac{u}{c}\right)^5$. The problem was first
addressed in 1941 by Lev D. Landau and Evgenij M. Lifshitz, in the
first edition of their textbook \cite{Landau}, where energy loss
was computed by applying the quadrupole formula. But it was not
clear whether their calculation was correct. At the time of the
Chapel Hill conference, there still was no consensus about the
result of such a computation, and not even on the fact that
gravitational waves were radiated at all. Differently from the
issue of the existence and detectability of gravitational waves,
this one was not settled at Chapel Hill, but continued to be
controversial for many years
\cite{Kennefick:2007zz}\footnote{Interestingly, one of the most
common objections towards the existence of a gravitational
radiation reaction was the lack of any evidence that advanced
potentials did not play any role in the theory. Some people
believed that maybe gravity worked like a Wheeler-Feynman absorber
theory \cite{FeynmanWheeler1}-\cite{FeynmanWheeler3}, in which the
relevant solution was a combination of retarded plus advanced
potential (even though the nonlinearity of the theory prevented a
combination of two solutions to be a rigorous solution), but
unlike the case of electromagnetism, the weakness of the force
prevented the existence of absorbers capable of breaking
time-reversal symmetry. Apparently, despite having developed the
absorber theory for electromagnetism, Feynman did not believe in
such ideas, since, as we shall see, he solved the wave equation
within linearized gravity in full analogy with standard
electrodynamics.}.

These issues captured the interest at the Chapel Hill Conference,
as clearly recalled by Peter G. Bergmann in his Summary
\cite{BergmannRMP} (p. 353):
\begin{quote}
In view of our interest in the role that gravitation, and
particularly its quantum properties, may play in microphysics, the
existence and the properties of gravitational waves represent an
issue of preeminent physical significance.
\end{quote}
Let us now turn to Feynman's contributions to both the issues of
the reality of gravitational waves, and of gravitational radiation
by binary star systems.

\subsection{The sticky-bead argument}

The question of the actual existence of gravitational waves
appeared in a discussion following one of Feynman's thought
experiments, in a session devoted to discussing the necessity of
gravity quantization (these thought experiments will be described
in Section 4). Feynman arrived one day late at the conference and
hence had not attended the session on gravitational waves.
Although he was trying to argue for the need to quantize the
gravitational field even in the absence of gravitational waves,
Feynman came out with a simple physical argument in favor of their
existence (\cite{ChapelHill}, p. 260 and pp. 279-281). He argued
that gravitational waves, if they exist, must carry energy and,
along with their existence as solutions of the gravitational
equations (albeit in the linear approximation), this was enough
for him to be confident in the possibility of their actual
generation. In his words: ``My instincts are that if you can feel
it, you can make it" (\cite{ChapelHill}, p. 260). Feynman reasoned
as follows:
\begin{quote}
Suppose we have a transverse-transverse wave generated by
impinging on two masses close together. Let one mass $A$ carry a
stick which runs past touching the other $B$. I think I can show
that the second in accelerating up and down will rub the stick,
and therefore by friction make heat. I use coordinates physically
natural to $A$, that is so at $A$ there is flat space and no field
(\cite{ChapelHill}, p. 279)\footnote{Feynman's words here are not
very precise, since it seems that the same two masses both
generate and receive the wave. Presumably, he was considering a
wave generated by two colliding masses and another couple of
masses (the ``beads'') as detector. It may also be that the words
``generated by'' have been transcribed by mistake, so that it is
the wave which impinges on the masses. However, the process which
generates the wave is not really relevant for the subsequent
argument, which instead is very clear.}.
\end{quote}
Then he recalled the result by Pirani \cite{Pirani1, Pirani2}
(which was presented earlier at the conference, cf.
\cite{ChapelHill}, p. 141), stating that the displacement $\eta$
of the mass $B$ (measured from the origin $A$ of the coordinate
system) in the field of a gravitational wave satisfies the
following differential equation:\footnote{This follows from the
geodesic deviation equation (\cite{ChapelHill}, p. 141;
\cite{Pirani1,Pirani2}).}
\begin{eqnarray}\label{PiraniEq}
\ddot{\eta}^a+R^{a}_{0b0}\eta^b=0, \qquad (a,b=1,2,3)
\end{eqnarray}
$R$ being the curvature tensor at $A$. The curvature does not
vanish for the transverse-transverse gravity wave but oscillates
as the wave goes by, so that Eq. (\ref{PiraniEq}) predicts that
the particle vibrates a little up and down, hence it rubs the
stick, producing heat by friction. This means that the stick
absorbs energy from the gravitational wave, leading to the
conclusion that gravitational waves carry energy.

Within the discussion (\cite{ChapelHill}, p. 260), Feynman also
quoted a result concerning the calculation of the energy radiated
by a two-body system (e.g. a binary star) in a circular orbit,
showing that he had addressed also the second issue mentioned
above, with detailed calculations:
\begin{eqnarray}\label{doublestar}
\frac{\text{Energy radiated in one revolution}}{\text{Kinetic
energy content}}=\frac{16 \pi}{15} \frac{\sqrt{mM}}{m+M} \left(
\frac{u}{c} \right)^5.
\end{eqnarray}
Here $m$ and $M$ are the masses of the two bodies, while $u$ is
the magnitude of their relative velocity. For the Earth-Sun
system, this formula describes a tiny effect, leading to a huge
order of magnitude (about $10^{26}$ years) for the lifetime of the
motion of the Earth around the Sun. The above argument was based
on a detailed analysis performed by Feynman some time before
(\cite{ChapelHill}, p. 280), but the full calculation (with some
differences, according to Feynman) was reported only four years
later in a letter written to Victor F. Weisskopf
\cite{WeisskopfLetter} (see next Subsection).

Hermann Bondi and Weber also attended the Chapel Hill conference.
Bondi likely envisaged an argument similar to
Feynman's\footnote{This is proved by several of his remarks at the
conference, such as the following one, made after Pirani's talk
(\cite{ChapelHill}, p.142): ``Can one construct in this way an
absorber for gravitational energy by inserting a $d\eta/d\tau$
term, to learn what part of the Riemann tensor would be the energy
producing one, because it is that part that we want to isolate to
study gravitational waves?''. The absorber of gravitational energy
is nothing but the ``stickiness'' of Feynman's sticky-bead.}, and
shortly after the conference he published a variant of the
sticky-bead device \cite{Bondi}, although he did not succeed in
relating the intensity of the gravitational wave to the amount of
the energy carried by it. The fact that gravitational waves were
real and physical was rigorously proved, in the framework of full
nonlinear general relativity, not much later, by Bondi himself,
Pirani, Ivor Robinson and Andrzej Trautman, who found exact
solutions describing plane waves \cite{Bondi:1958aj}, and waves
radiated from bounded sources \cite{Robinson:1960zzb}. In the
sixties, Bondi, Rainer W. Sachs, Ezra T. Newman and Roger Penrose
\cite{Bondi:1962px} - \cite{Penrose:1965am}, gave a satisfactory
definition of the energy radiated at infinity as gravitational
radiation by an isolated gravitating system. Weber soon began to
study how to experimentally detect gravitational waves
\cite{Weber1}, building his famous resonant bar detector in 1966
\cite{Weber2} and announcing the first experimental results (soon
to be disproved) in 1969 \cite{Weber3}. These works triggered all
the subsequent research on the {\it detection} of gravitational
waves.

\subsection{Further elaboration}

Although Feynman never published anything on gravitational waves,
a complete and more systematic description of his Chapel Hill
proposal can be found in the already mentioned letter he wrote to
Weisskopf in February 1961 \cite{WeisskopfLetter}, and
subsequently included in the material distributed to the students
attending his Caltech lectures \cite{Feynman:1996kb} in
1962-63\footnote{In fact, the computations in the letter must have
been somewhat simpler, since Feynman admits that ``Only as I was
writing this letter to you did I find this simpler argument''
(\cite{WeisskopfLetter}, p.14).}. A part of that calculation can
in fact also be found in Lecture 16. Feynman indeed claimed that
\begin{quote}
It was this entire argument used in reverse that I made at a
conference in North Carolina several years ago to convince people
that gravity waves must carry energy.  (\cite{WeisskopfLetter}, p.
14).
\end{quote}
Feynman's reasoning developed in close analogy with
electrodynamics, but taking into account the fact that in the case
of gravitation the source is a tensor $S_{\mu\nu}$ rather than a
vector $A_{\mu}$, so that the following differential equation has
to be solved to find classical gravitational waves:
\begin{eqnarray}\label{waves1}
\Box^2 \overline{h}_{\mu\nu}=\lambda S_{\mu\nu}.
\end{eqnarray}
Here as usual $\overline{h}_{\mu\nu}$ represents the  metric
perturbation (the bar operation is defined in Eq.
(\ref{barOperation}) below), $\Box^2$ is the usual, flat-space,
d'Alembertian operator and the de Donder gauge
$\partial^{\mu}\overline{h}_{\mu\nu}=0$ is used. When all
quantities are periodic with frequency $\omega$, the solution at a
point $1$, which is located at a distance much greater from the
source (i.e. the region where $S_{\mu\nu}$ is expected to be
large) than the dimensions of the source itself, is:
\begin{eqnarray}\label{SolWaves1}
\overline{h}_{\mu\nu}(\vec r_1)=- \frac{\lambda}{4\pi r_1} e^{i
\omega r_1} \int d^3 \vec r_2 S_{\mu\nu}(\vec r_2) e^{-i \vec K
\cdot \vec r_2},
\end{eqnarray}
with $|\vec r_1|\gg |\vec r_2|$. Thus the first non-vanishing term
in the sequence of integrals corresponding to an expansion of the
exponential is a quadrupole one. This approximation holds for
nearly all cases of astronomical interests, where wavelengths are
much longer than the system's dimensions, such as for instance
double stars or the earth-sun system. Feynman's attention then
concentrated on computing the power radiated by the above waves in
the quadrupole approximation which, for a periodic motion of
frequency $\omega$, reads (\cite{WeisskopfLetter}, p.11):
\begin{eqnarray}\label{PowerRad}
\frac{1}{5} G \omega^6 \sum_{ij} \left| Q_{ij}^{'} \right|^2,
\end{eqnarray}
where $Q_{ij}^{'}=Q_{ij}-\frac{1}{3} \delta_{ij} Q_{kk}$,
$Q_{ij}=\sum_{a} m_a R_i^a R_j^a$ being the quadrupole moment of
mass, and the RMS average is taken. In particular, for a
circularly rotating double star, the result (\ref{doublestar}) is
recovered. In sum, Feynman applied the quadrupole formula to a
binary system, as Landau and Lifshitz had earlier. Feynman's
treatment was simpler, however, since he managed by a clever trick
to avoid referring to the energy-momentum pseudo-tensor (this is
the ``simpler argument'' referred to above). The question of
course arises, whether he took inspiration from the book by Landau
and Lifshitz, or rather devised his approach in a completely
independent fashion\footnote{The issue of whether Feynman took
inspiration from Landau-Lifshitz book also arises concerning his
formulation of Maxwell's equations, and was briefly considered by
the present authors in \cite{DiMauro:2020bpd}, without reaching a
definite conclusion.}.

The above treatment is supplemented with a thorough analysis of
the effects of a gravitational wave impinging on a device made of
two test particles placed on a rod with friction, in this way
providing a more complete description of the working principle of
the gravitational wave detector previously introduced at Chapel
Hill. A second absorber, made of four moving particles in a
quadrupole configuration, is also proposed, and shown to be able
to absorb energy from a gravity wave acting on it. The same
oscillating device is shown to re-radiate waves with the same
energy content. Apparently, at Chapel Hill, Feynman had derived
the expression for the radiated energy from the detailed study of
the detector, while in the letter the opposite route is taken
(hence he says that the argument is ``used in reverse'').

These results went beyond the treatment carried out in the last
section of Lecture 16 of the Caltech Lectures
(\cite{Feynman:1996kb}, pp. 218-220), where the treatment stops
immediately after the solution of the wave equation. An
interesting point is that, both in the letter and in the Lectures,
Feynman performed the computation twice, first using the methods
of quantum field theory to compute the tree level probability for
low-energy graviton emission in several scattering processes, and
then by solving the classical linearized wave equation
(\ref{waves1}) to compute the intensity of the emitted waves, of
course with agreeing results (in fact the tree approximation of a
quantum theory is equivalent to its classical limit). At the end
of Lecture 16 of \cite{Feynman:1996kb}, Feynman commented again on
the energy content of gravitational waves:
\begin{quote}
We can definitely show that they can indeed heat up a wall, so
there is no question as to their energy content. The situation is
exactly analogous to electrodynamics (\cite{Feynman:1996kb}, p. 219).
\end{quote}
Interestingly, in the letter, while remembering the Chapel Hill
conference he comments:
\begin{quote}
I was surprised to find a whole day at the conference devoted to
this question, and that ``experts'' were confused. That is what
comes from looking for conserved energy tensors, etc. instead of
asking  ``can the waves do work?'' (\cite{WeisskopfLetter}, p.
14).
\end{quote}
Indeed, from what he wrote in the letter it is clear that he
considered the issue of whether binary systems radiate as solved
since his calculations clearly showed that the phenomenon
occurred. Indeed the fact that relativists continued to argue
about the binary star problem without reaching an agreement
disturbed Feynman, and may have contributed to his loss of
interest in the subject.

\section{Should gravity be quantized?}

One of the most debated questions at the Chapel Hill conference
was whether the gravitational field had to be quantized at all.
This was a crucial issue in view of the merging of general
relativity with the rest of physics, especially with the then
thriving field of elementary particle physics, which was of course
dominated by quantum mechanics. As remarked by Bergmann
(\cite{ChapelHill}, p. 165):
\begin{quote}
Physical nature is an organic whole, and various parts of
physical theory must not be expected to endure in ``peaceful
coexistence.'' An attempt should be made to force separate
branches of theory together to see if they can be made to merge,
and if they cannot be united, to try to understand why they clash.
Furthermore, a study should be made of the extent to which
arguments based on the uncertainty principle force one to the
conclusion that the gravitational field must be subject to quantum
laws: (a) Can quantized elementary particles serve as sources for
a classical field? (b) If the metric is unquantized, would this
not in principle allow a precise determination of both the
positions and velocities of the Schwarzschild singularities of
these particles?
\end{quote}
Besides the many technical talks and comments, where the main
roads to quantum gravity which would have been pursued in the
following decades were delineated, much effort was devoted to
fundamental and conceptual questions, in view of the feeling that
physicists should
\begin{quote}
attempt to keep physical concepts as much as possible in the
foreground in a subject which can otherwise be quickly flooded by
masses of detail and which suffers from lack of experimental
guideposts (\cite{ChapelHill}, p. 167).
\end{quote}
Hence, the problem of quantum measurement was discussed at length
during the conference. As stated by Bergmann, the main conceptual
question was:
\begin{quote}
What are the limitations imposed by the quantum theory on the
measurements of space-time distances and curvature?
(\cite{ChapelHill}, p. 167)
\end{quote}
or, equivalently
\begin{quote}
What are the quantum limitations imposed on the measurement of the
gravitational mass of a material body, and, in particular, can the
principle of equivalence be extended to elementary particles?
(\cite{ChapelHill}, p. 167)
\end{quote}
The editors of Ref. \cite{ChapelHill} pointed out that the answer
could not be simply given in terms of dimensional arguments, since
the Planck mass does not constitute a lower limit to the mass
of a particle whose gravitational field can in principle be
measured. A simple argument (\cite{ChapelHill}, pp. 167-8) shows
in fact that the gravitational field of \emph{any} mass can in
principle be measured, thanks to the ``long tail'' of the
Newtonian force law.

\subsection{Unavoidable quantum effects}

Feynman's views concerning such general matters were very close to
those of Bergmann (while they disagreed on more technical ones).
Indeed, as his interventions in the many discussion sessions
clearly show, Feynman was convinced that nature cannot be half
classical and half quantum, as he opposed those proposing that
maybe gravity should not be quantized. Feynman's arguments in
favor of the quantization of the gravitational field were quite
different from the ones that became popular later among most
practitioners, involving the assumption that at the Planck scale
gravity has to dominate over all other interactions. He thought
that it was not necessary to go to such ridiculously high and
unfathomable scales to see quantum effects involving gravity.
Specifically, he proposed a thought experiment showing that,
assuming the validity of quantum mechanics for objects massive
enough to produce a detectable gravitational field (an assumption
in which he definitely believed, since the opposite would have
required a modification of quantum mechanics \cite{Zeh}), then the
only way to avoid contradictions is to quantize the gravitational
field itself (\cite{ChapelHill}, pp. 250-252). Alongside with the
consideration that a mass point giving rise to a classical
Schwarzschild field very unlikely would obey Heisenberg's
uncertainty relations, a more pragmatic motivation was the hope
(also expressed earlier by Wolfgang E. Pauli) that quantization of
the metric would help resolving the divergencies present in
quantum field theory and, in this way, it would have a key
relevance also for the theory of elementary particles. This hope
underlies, for example, Feynman's criticism to Deser's proposal,
where all non-gravitational fields should be quantized in a given
geometry, while the quantization of the metric field itself would
be the final step of the whole procedure \cite{BergmannRMP}.
Quoting Feynman:
\begin{quote}
If one started to compute the mass correction to, say, an
electron, one has two propagators multiplying one another each of
which goes as $1/s^2$ and are singular for any value of the $g$'s.
Therefore, do the spatial integrations first for a fixed value of
the $g$'s and the propagator is singular, giving $\delta
m=\infty$. Then the superposition of various values of the $g$'s
is still infinite (\cite{ChapelHill}, p. 270).
\end{quote}
Feynman's approach to quantum gravity, as we shall see, was
instead fully quantum from the beginning.

\subsection{Toward quantum gravity}

The above-mentioned discussions were one of the few places where
Feynman was directly involved in the foundations and
interpretation of quantum mechanics \cite{Zeh} or, at least, where
his thoughts about the subject were expressed. His arguments,
which were based essentially on thought experiments, stimulated a
wide debate on the measurement problem in quantum mechanics and on
the existence and meaning of macroscopic quantum superpositions
\cite{ChapelHill}. The debate further grew in Session VIII of the
conference, where the discussion focused on the contradictions
eventually arising in the logical structure of quantum theory if
quantization of gravity was not assumed. Indeed, as pointed out by
DeWitt (\cite{ChapelHill}, p. 244), the choice of the quantum
expectation value of the stress-energy tensor of matter fields as
the source of the gravitational field may lead to difficulties,
since measurements made on the system may change this expectation
value and hence the gravitational field.

The validity of the classical theory of gravitation relies on the
smallness of fluctuations at the scale where gravitational effects
become sizeable. Feynman's first argument concerned a two-slit
diffraction experiment with a mass indicator behind the two-slit
wall, i.e. a gravitational two-slit experiment. If one is working
in a space-time region whose linear dimensions are of order $L$ in
space and ${L}/{c}$ in time, the uncertainty on the gravitational
potential (divided by $c^2$ so that it is dimensionless and thus
homogeneous to a metric) is in general $\Delta
g=\sqrt{\frac{hG}{c^3L^2}}=\frac{L_P}{L}$, where
$L_P=\sqrt{\frac{hG}{c^3}}$ is the Planck length (as argued for
example by Wheeler in \cite{ChapelHill}, pp. 179-180, on the basis
of the path integral for the gravitational field). The order of
magnitude of the potential generated by a mass $M$ within the
spatial part of the considered region is $g=\frac{MG}{Lc^2}$,
hence in this case $\Delta g=\frac{\Delta M G}{Lc^2}$, from which
by comparison with the general expression one could infer a mass
uncertainty $\Delta M =\frac{c^2 L_P}{G}\approx 10^{-5}$ grams, if
the time of observations is less than ${L}/{c}$. On the other
side, by allowing for an infinite time, one would obtain $M$ with
infinite accuracy. Feynman deduced that the last option could not
take place for a mass $M$ put into the above two-slit apparatus,
so that the apparatus would not be able to uncover this
difficulty, unless $M$ was at least of order $10^{-5}$ grams. His
conclusion was that:
\begin{quote}
Either gravity must be quantized because a logical difficulty
would arise if one did the experiment with a mass of order
$10^{-4}$ grams, or else [...] quantum mechanics fails with masses
as big as $10^{-5}$ grams (\cite{ChapelHill}, p. 245).
\end{quote}
The discussion then briefly dealt with some formal matters and
with the meaning of the equivalence principle in quantum gravity,
with Helmut Salecker proposing a thought experiment which
apparently lead to a violation of the equivalence principle in the
quantum realm. Finally, due to a remark by Salecker himself, the
discussion again focused on the topical question. Salecker pointed
out that charged quantized particles might act as a source of an
unquantized Coulomb field within an action-at-a-distance picture,
perhaps hinting (as commented by an Editor's Note,
\cite{ChapelHill}, p. 249) to a similar situation in the case of a
gravitational field. Subsequent considerations by Frederick J.
Belinfante suggested the quantization of the static part of the
gravitational field as well as of the transverse part (which
describes the gravitational radiation), in order to avoid
difficulties arising from the choice of an expectation value (of
the stress-energy tensor) as the source of the gravitational
field, in full agreement with the previous argument by DeWitt.
Here, as noted by H. Dieter Zeh \cite{Zeh}, Belinfante's
understanding of the wave function as an epistemic concept clearly
emerged, since he said:
\begin{quote}
``There are two quantities which are involved in the description
of any quantized physical system. One of them gives information
about the general dynamical behavior of the system, and is
represented by a certain operator (or operators). The other gives
information about our knowledge of the system; it is the state
vector [...] the state vector can undergo a sudden change if one
makes an experiment on the system''\footnote{This expression
reveals that according to Belinfante ``the wave function [...]
must change for reasons beyond the system's physical dynamics. He
[Belinfante] does not refer to ensembles of wave functions or a
density matrix in order to represent incomplete knowledge''
(\cite{Zeh}, p. 65).} (\cite{ChapelHill}, p. 250).
\end{quote}
Feynman replied with his second thought experiment, a
Stern-Gerlach experiment with a gravitational apparatus, which
revealed his completely different ideas on the meaning of
quantization and on the role of wave function. Feynman's argument
dealt with a spin-$1/2$ particle going through the apparatus and
then crossing one of two counters (denoted as $1$ and $2$,
respectively), each one connected by means of a rod to an
indicator, which is a little ball with a diameter of $1$ cm, going
up or down depending on whether the object arrives at counter $1$
or $2$, respectively. A quantum mechanical analysis provides, in
principle (before making a measurement), an amplitude for the ball
up and an amplitude for the ball down\footnote{As pointed out by
Zeh \cite{Zeh}, Feynman's description of the measurement process
is very close to the standard measurement and registration device
proposal by von Neumann \cite{VonN}.}. Thanks to its macroscopic
size, the ball is able to produce a gravitational field and such a
field may be used to move a probe ball. Thus the gravitational
field acts as a channel between the object and the observer. This
reasoning leads Feynman to the following conclusion about gravity
quantization:
\begin{quote}
Therefore, there must be an amplitude for the gravitational field,
provided that the amplification necessary to reach a mass which
can produce a gravitational field big enough to serve as a link in
the chain does not destroy the possibility of keeping quantum
mechanics all the way. There is a bare possibility (which I
shouldn't mention!) that quantum mechanics fails and becomes
classical again when the amplification gets far enough, because of
some minimum amplification which you can get across such a chain.
But aside from that possibility, if you believe in quantum
mechanics up to any level then you have to believe in
gravitational quantization in order to describe this experiment
(\cite{ChapelHill}, p. 251).
\end{quote}
To a question by Bondi, who asks (\cite{ChapelHill}, p. 252):
``What is the difference between this and people playing dice, so
that the ball goes one way or the other according to whether they
throw a six or not?'', Feynman answers:
\begin{quote}
I don't really have to measure whether the particle is here or
there. I can do something else: I can put an inverse Stern-Gerlach
experiment on and bring the beams back together again. And if I do
it with great precision, then I arrive at a situation which is not
derivable simply from the information that there is a 50 percent
probability of being here and a 50 percent probability of being
there. In other words, the situation at this stage is not 50-50
that the die is up or down, but \emph{there is an amplitude} that
it is up and an amplitude that it is down -- a complex amplitude
-- and as long as it is still possible to put those amplitudes
together for interference you have to keep quantum mechanics in
the picture (\cite{ChapelHill}, p. 252, our emphasis).
\end{quote}
As Zeh notices (\cite{Zeh} p. 67), this is a standard argument
against an epistemic interpretation of the wave function.

\subsection{Wave packet reduction}

In the last sentence of the above quote, Feynman focused on the
problem of wave function collapse and hints to decoherence as a
possible solution. He argued that the wave packet reduction acts
somewhere in his experimental apparatus thanks to the amplifying
mechanism, so that amplitudes become probabilities in the presence
of a huge amount of amplification (via the macroscopic
gravitational field of the ball). Then he wondered whether it
might be possible to design an experiment in which the wave packet
reduction due to the amplification process could be avoided.
Subsequent critical remarks by Leon Rosenfeld and Bondi lead
Feynman to envision a kind of quantum interference in his
experiment, by allowing gravitational interaction between
macroscopic balls to be described by means of a quantum field with
suitable amplitudes taking a value or another value, or to
propagate here and there. Clearly, as suggested by Bondi, one has
to remove any irreversible element such as, for instance, the
possibility that gravitational links radiate. This is probably
another hint to the possible role of dissipation or decoherence in
destroying quantum interference, but Feynman and Bondi here only
speak about classical irreversibility \cite{Zeh}. The concept of
decoherence as source of smearing off phase relations, as well as
the transition to the classical picture and its environmental
origin, was in fact still unclear at the time (for a historical
and research account see Ref. \cite{ZurekRev} and references
therein). By arguing again on gravity quantization in relation to
wave function collapse, Feynman claimed that:
\begin{quote}
There would be a \textit{new principle}! It would be
\textit{fundamental}! The principle would be: -- \textit{roughly}:
\textit{Any piece of equipment able to amplify by such and such a
factor} ($10^{-5}$ grams or whatever it is) \textit{necessarily
must be of such a nature that it is irreversible}. It might be
true! But at least it would be fundamental because it would be a
new principle. There are two possibilities. Either this principle
-- this missing principle -- is right, \textit{or} you can amplify
to any level and still maintain interference, in which case it's
absolutely imperative that the gravitational field be quantized...
\textit{I believe}! \textit{or} there's another possibility which
I haven't thought of (\cite{ChapelHill}, pp. 254-255, emphasis in
original).
\end{quote}
The discussion on possible sources of decoherence went on further,
with Feynman assuming that quantum interference might eventually
take place with a mass of macroscopic size, i.e. about $10^{-5}$
grams or even $1$ gram, and hinting to the possible role of
gravity in destroying quantum superpositions. The same argument
would be later developed by Feynman in his Lectures on Gravitation
\cite{Feynman:1996kb}, where he dealt with ``philosophical
problems in quantizing macroscopic objects'' and commented about a
possible gravity-induced failure of quantum mechanics:
\begin{quote}
I would like to suggest that it is possible that quantum
mechanics fails at large distances and for large objects. Now,
mind you, I do not say that I think that quantum mechanics
\textit{does} fail at large distances, I only say that it is not
inconsistent with what we do know. If this failure of quantum
mechanics is connected with gravity, we might speculatively expect
this to happen for masses such that $\frac{GM^2}{\hbar c}=1$, of
$M$ near $10^{-5}$ grams, which corresponds to some $10^{18}$
particles (\cite{Feynman:1996kb}, pp. 12-13).
\end{quote}
The same problem was pointed out again later
(\cite{Feynman:1996kb}, p. 14), where the possibility was put
forward that amplitudes may reduce to probabilities for a
sufficiently complex object, as a consequence of a smearing effect
on the evolution of the phases of all parts of the object. Such a
smearing effect could be related to the existence of gravitation.
A similar idea was expressed in the end of the letter to Weisskopf
\cite{WeisskopfLetter}, where he wrote:
\begin{quote}
How can we experimentally verify that these waves are quantized?
Maybe they are not. Maybe gravity is a way that quantum mechanics
fails at large distances.
\end{quote}
It is remarkable that, although Feynman often expressed his belief
in the quantumness of nature, including gravity, he was
nevertheless open to the possibility that gravity may not be
quantized, in the absence of experiments able to clear the question.

\subsection{Feynman-inspired later results}

The possibility of a gravity-induced collapse of the wave
function, as a solution to the measurement problem in quantum
mechanics, can thus be traced back to the broad debate at Chapel
Hill about gravity quantization and Feynman's insights. This is an
attractive idea, because gravity is ubiquitous in all existing
interactions, and gravitational effects depend on the size of
objects, and it triggered a huge amount of investigation (see for
example \cite{karol}-\cite{penrose3}). In particular, the Penrose
proposal relies strongly on the conflict between the general
covariance of general relativity and the quantum mechanical
superposition principle \cite{penrose0,penrose1}: such a conflict
emerges when considering a balanced superposition of two separate
wave packets representing two different positions of a massive
object. If the mass $M$ of this object is large enough, the two
wave packets represent two very different mass distributions. By
assuming that in each space-time one can use the notions of
stationarity and energy, while the difference between the
time-translation operators gives a measure of the ill-definiteness
or uncertainty of the final superposition's energy, the decay time
for the balanced superposition of two mass distributions is:
\begin{eqnarray}\label{PenroseEnergy}
t_D=\frac{\hbar}{\Delta E_{grav}}.
\end{eqnarray}
Here $\Delta E_{grav}$ is the gravitational self-energy of the
difference between the mass distributions of each of the two
locations of the object. Thus, massive superpositions cannot form,
because they would decay immediately. Later, Penrose suggested
that the basic stationary states into which a superposition of
such states decays are stationary solutions of the so-called
Newton-Schr\"{o}dinger equation \cite{penrose2}. More recently,
other collapse models (called \textit{dynamical} collapse models)
have been also suggested \cite{random1}-\cite{random4}, where the
collapse of the wave function is induced by the interaction with a
random source, for instance an external noise source. In the
gravitational context, a lot of experimental proposals have been
put forward as well, attempting to explore a parameter regime
where both quantum mechanics and gravity are significant.
Currently, it has been possible to build up a quantum
superposition state with complex organic molecules with masses of
the order $m=10^{-22}$ kg \cite{arndt1,arndt2}. Today prospective
typical experiments, revealing gravity-induced decoherence,
involve matter-wave interferometers
\cite{matterW1}-\cite{matterW3}, quantum optomechanics
\cite{bose}-\cite{oriol} and magnetomechanics \cite{mag1}.
However, despite the huge theoretical and experimental effort,
there is still no consensus on a definitive solution of the
fundamental dichotomy between unitary deterministic quantum
dynamics and the discontinuous irreversible state collapse
following a measurement process, as well as of the problem of the
emergence of classical from the quantum world.

\subsection{Observers in a closed Universe}\label{Everett}

Another foundational topic which emerged from discussions on
quantum gravity is the role of the observer in a closed Universe.
As pointed out by Wheeler:
\begin{quote}
General relativity, however, includes the space as an integral part of the
physics and it is impossible to get outside of space to observe the physics.
Another important thought is that the concept of eigenstates of the total
energy is meaningless for a closed Universe. However, there exists the
proposal that there is one ``universal wave function". This function has
already been discussed by Everett, and it might be easier to look for this
``universal wave function" than to look for all the propagators (\cite{ChapelHill}, p. 270).
\end{quote}
As already stated in Section 2.1, Wheeler was giving the first
presentation of Everett's relative-state interpretation of quantum
mechanics \cite{Everett,Osnaghi}, to which Feynman promptly
replied with a ``many-worlds" characterization:
\begin{quote}
The concept of a ``universal wave function" has serious conceptual
difficulties. This is so since this function must contain amplitudes for all
possible worlds depending on all quantum-mechanical possibilities in the
past and thus one is forced to believe in the equal reality of an infinity
of possible worlds (\cite{ChapelHill}, p. 270).
\end{quote}
The same idea would have been expressed by Feynman some years
later in his Lectures on Gravitation (\cite{Feynman:1996kb}, pp.
13-14), where the role of the observer in quantum mechanics was
also discussed by making explicit reference to the Schr\"{o}dinger
cat paradox. In particular, an external observer is in a peculiar
position because he always describes the result of a measurement
by an amplitude, while the system collapses into a well-defined
final state after the measurement. On the other hand, according to
an internal observer, the result of the same measurement is given
by a probability. Thus, a paradoxical situation emerges in the
absence of an external observer, most notably in considering the
whole Universe as being described by a complete wave function
without an outside observer. The Universe wave function obeys a
Schr\"{o}dinger equation and implies the presence of an infinite
number of amplitudes, which bifurcate from each atomic
event\footnote{Here Feynman probably hints to a vision of the
Universe as constantly splitting into an infinite number of
branches, which result from the interactions between its
components. Indeed such interactions act as measurements.}. This
implies that an inside observer knows which branch the world has
taken, so that he can follow the track of his past. Feynman
concluded the argument by raising a conceptual problem:
\begin{quote}
Now, the philosophical question before us is, when we make an
observation of our track in the past, does the result of our
observation become real in the same sense that the final state
would be defined if an outside observer were to make the
observation (\cite{Feynman:1996kb}, p. 14)?
\end{quote}
Later in the Lectures on Gravitation (\cite{Feynman:1996kb}, pp.
21-22), Feynman returned briefly to the meaning of the wave
function of the Universe and confirmed his ``many-worlds"
characterization of Everett's approach by resorting to a ``cat
paradox on a large scale'', from which our world could be obtained
by a ``reduction of the wave packet". He questioned the mechanism
of this reduction, the crucial issue being how to relate Everett's
approach and collapse mechanisms of whatever origin. In this
connection, it is interesting to quote a comment made by John
Preskill in a recent talk about the Feynman legacy (see
\cite{PreskillTalk}, slide 29):
\begin{quote}
When pressed, Feynman would support the Everett viewpoint, that
all phenomena (including measurement) are encompassed by unitary
evolution alone. According to Gell-Mann, both he and Feynman
already held this view by the early 1960s, without being aware of
Everett's work\footnote{We do not agree that Feynman was not aware
of Everett's work in the 1960s, since he commented on it at Chapel
Hill in 1957.}. However, in 1981 Feynman says of the many-worlds
picture: ``It's possible, but I am not very happy with it".
\end{quote}
Everett's analysis \cite{Everett} is, indeed, the first attempt to
go beyond the Copenhagen interpretation in order to apply quantum
mechanics to the Universe as a whole. This requires to overcome
the sharp separation of the world into ``observer" and
``observed", showing how an observer could become part of the
system while measuring, recording or doing whatever operation
according to the usual quantum rules. Quantum fluctuations of
space-time in the very early Universe have also to be properly
taken into account. On the other hand, this approach lacks an
adequate description of the origin of the quasi-classical realm as
well as a clear explanation of the meaning of the branching of the
wave function. A further extension and completion of Everett's
work has been developed by a number of authors
\cite{dh1}-\cite{dh7} and is today known as \textit{decoherent
histories approach to quantum mechanics of closed systems}. Within
this formulation neither observers nor their measurements play a
prominent role, and the so called \textit{retrodiction}, namely
the ability to construct a history of the evolution of the
Universe toward its actual state by using today's data and an
initial quantum state, is allowed. The process of prediction
requires to select out decoherent sets of histories of the
Universe as a closed system, while decoherence in this context
plays the same role of a measurement within the Copenhagen
interpretation. Decoherence is a much more observer-independent
concept and gives a clear meaning to Everett's branches, the main
issue being the identification of mechanisms responsible for it.

\section{Gravity as a quantum field theory}

In a set of critical comments at Chapel Hill (\cite{ChapelHill},
pp. 272-276), Feynman advocated a non-geometric and field
theoretical approach to gravity, which would later be developed in
the first part of his Caltech lectures \cite{Feynman:1996kb}. He
imagined a parallel development of history\footnote{In
\cite{ChapelHill}, p. 273, Feynman said: ``Instead of trying to
explain the rest of physics in terms of gravity I propose to
reverse the problem by changing history. Suppose Einstein never
existed [...]".}, in which the general theory of relativity had
not been discovered yet, while the principles of Lorentzian
quantum field theory were known. Then, he wondered, how would
people deal with the discovery of a new force, namely gravitation?
The roots of such an approach can be traced back to Robert H.
Kraichnan \cite{Kraichnan:1955zz} and Suraj N. Gupta
\cite{Gupta:1954zz} (more details about this fascinating story can
be found below, in Section 5.1, and in \cite{LectGravPreface}).
Basically, one can use general arguments from field theory and
from experiment to infer that gravity (assumed to be mediated by
virtual particle exchanges as any other interaction) has to be
carried by a massless neutral spin-2 quantum, called the graviton.
Then, full general relativity should follow from the the quantum
theory of a massless neutral spin-2 field\footnote{As we discuss
in more detail below, the linear theory for such a field and its
massive counterpart was completely worked out by Markus Fierz and
Pauli in Ref. \cite{Fierz:1939ix}, following a previous work by
Paul A. M. Dirac \cite{Dirac:1936tg}.}, as well as from
consistency requirements\footnote{See, however,
\cite{Padmanabhan:2004xk}.}. The same procedure, for the spin-1
case, yields the Maxwell equations. Such an approach is also
coherent with Feynman's views about fundamental interactions
\cite{DiMauro:2020bpd}.


\subsection{Quantum gravity research before 1957: a brief
account}

Before describing Feynman's work, let us briefly outline the
history of quantum gravity prior to the Chapel Hill conference.
Although the first embryonic ideas about the interaction of
quantum theory and gravity can be traced back at least to
Einstein's first paper on gravitational waves
\cite{Einstein:1916cc}, we here focus on the problem of quantizing
the gravitational field itself, which began to be seriously
tackled only after quantum mechanics was completed and the first
papers on field quantization were written. The interested reader
can refer to the book \cite{Rickles1}, as well as to the source
book \cite{RicklesBlum} for a comprehensive survey of early work
on quantum gravity\footnote{Interestingly enough, the exact words
``quantum gravity'' were first recorded only in 1969 (see e.g.
\cite{HartzBlum} and references therein).}, including both
pioneering work preceding 1930, and the part of the story which we
summarize here.

The first attempts to quantize the gravitational field date back
to the early 1930s, when the bases for the so-called covariant and
canonical approaches were laid. While the idea behind the
covariant approach is to build up a quantum field theory of the
perturbations $h_{\mu\nu}$ of the metric over a flat Minkowski
space, the canonical approach aims at recasting general relativity
in the Hamiltonian form (which requires singling out the time
direction, thus breaking explicit covariance), and then quantizing
it along the usual lines.

A mention of the quantization of the gravitational field in
analogy with the electromagnetic field can be found in the first
paper on quantum electrodynamics by Werner Heisenberg and Pauli
\cite{HeiPau1}. Their remarks prompted a subsequent attempt by
Rosenfeld \cite{Ros1,Ros2}\footnote{For Rosenfeld's earlier
incursions in quantum gravity, see \cite{Peruzzi:2018nzo}.}, who
applied a novel method for the quantization of fields with gauge
groups \cite{Salisbury1}-\cite{Salisbury:2009cr}, the Hamiltonian
formulation of which involves constraints, to the linearized
Einstein field equations. In the process, he showed that the
gravitational field has to be quantized by using commutators,
hence its quanta obey Bose statistics. Despite not completing the
program of quantizing linearized general relativity, in his second
paper Rosenfeld calculated the gravitational self-energy of a
light quantum (finding a divergence) and studied various possible
transition processes involving both light quanta and gravitational
quanta. But it is worth mentioning that, at the beginning of the
1930s, Rosenfeld did not realize yet that certain peculiar
features of gravity\footnote{We refer in particular to the
existence of gravitational collapse and the deep link of gravity
with spacetime structure, whose influence on the problem of
quantization began to be widely appreciated only in the fifties,
although it was recognized earlier by a few physicists, as we
discuss below.} would have posed fundamental problems in
quantizing the gravitational field. The idea of a gravitational
quantum analogous to the photon emerged in the same years (the
1930s) \cite{Stachel}, and the name graviton came out for the
first time in a 1934 paper by Dmitri I. Blokhintsev and F\"{e}dor
M. Gal'perin \cite{Galperin}, although those authors supposed that
the gravitational quantum was related to the
neutrino\footnote{Such an idea originated from Bohr, and was
discussed, among others, also by Pauli and Enrico Fermi (see
\cite{Stachel} and references therein). It was briefly considered
also by Feynman in his lectures \cite{Feynman:1996kb} (see below,
Section \ref{Neutrino}), only to show its untenability.}. Later,
in 1939, Fierz and Pauli \cite{Fierz:1939ix}, while studying the
quantization of fields with arbitrary spin and mass (and of their
coupling with the electromagnetic field), realized that a massless
spin-2 quantum field obeys equations which formally coincide with
the linearized Einstein equations, unveiling an interesting link
with Rosenfeld's work \cite{Ros1,Ros2}, and at the same time
showing that linearized quantum gravity can be developed from
purely quantum field theoretical considerations, independently of
general relativity. Meanwhile, in 1935, the quantization of the
linearized theory had been completely carried out independently by
Matvei P. Bronstein
\cite{Bronstein,Gorelik:2005an}\footnote{Interestingly, in the
course of this work Bronstein derived the quadrupole formula by
computing the emission coefficients of transverse gravitons, with
the correct factor of 2 that had been found by Eddington
\cite{Eddington:1922ds}. This computation is conceptually
analogous to the quantum computation performed by Feynman (by
means of more modern tools) in his letter to Weisskopf
\cite{WeisskopfLetter}. Since Bronstein's work was not very well
known in the West until the 1990s
\cite{GorelikBook,BronsteinComment}, it is likely that Feynman did
not know it. Moreover, Bronstein showed that Newton's law, with
the correct sign, is generated by the exchange of non-transverse
gravitons.} by means of Fermi's quantization technique
\cite{Fermi:1932xva} but, unlike Rosenfeld, he soon realized the
deep difference between quantum electrodynamics and a quantum
theory of gravitation. Bronstein carried out an analysis of the
measurements of the linearized Christoffel symbols (which he
identified with the components of the gravitational field),
modeled on the analogous one carried out by Niels Bohr and
Rosenfeld himself for the electromagnetic case
\cite{BohrRosenfeld}. The result was an expression for the minimum
uncertainty in a measurement which is a function of the inverse
mass density of a test body, so that this uncertainty cannot be
made arbitrarily small, since general relativity does not allow
the existence of arbitrarily massive bodies in a given volume. He
speculated that the validity of such a result would be preserved
also in the full nonlinear theory, and that the quantization of
the latter would imply a rejection of the usual concepts of space
and time. Bronstein's latter argument was criticized by Jacques
Solomon \cite{Solomon}, who however put forward his own argument
for the same conclusion, by questioning that a quantization method
based on the superposition principle would work when the
gravitational field is not weak. Interestingly, Solomon's argument
was based on a recent ``proof'' by Rosen \cite{Rosen} of the
nonexistence of non-singular plane gravitational waves in full
general relativity (the standard quantization methods that were
used at the time were in fact based on expansions of the fields in
plane waves). Of course, Rosen's proof was flawed, since what he
found were mere coordinate singularities (there is an interesting
analogy with the Einstein-Rosen cylindrical waves
\cite{EinsteinRosen1937}). In fact, exact plane wave solutions
were found in the late 1950s \cite{Bondi:1958aj}, as we discussed
in Section 3.2.

The quantum gravity landscape began to change in the late 1940s
\cite{HartzBlum}. After the great successes of renormalized
quantum electrodynamics, the idea that full quantum general
relativity could be treated by means of a perturbative expansion
around flat space, with nonlinear terms seen as self-interactions
of the gravitational field with itself, gained momentum. DeWitt
was the first who pursued such an approach in his Ph.D. thesis
\cite{DeWittPHD}, applying the powerful covariant machinery
developed by his advisor Julian S. Schwinger \cite{Schwinger} to
the computation of the photon gravitational self-energy. Unlike
Rosenfeld, he found a vanishing result, in agreement with the
requirements of gauge invariance. The already mentioned program of
constructing the \emph{full} theory of general relativity by
starting from a free, massless spin-2 field in Minkowski space,
which is closely related to this approach to quantization, and
which was embraced also by Feynman, started being developed by
Kraichnan \cite{Kraichnan:1955zz} (who actually began pursuing it
back in 1947) as well as Gupta \cite{Gupta:1954zz} (for further
details see \cite{LectGravPreface,HartzBlum} and a comment in the
following subsection). In the same years, another line of research
became popular among relativists, who were just starting to
recognize nonlinearity as the essential feature of general
relativity (this was in fact one of the aspects that characterized
the renaissance of general relativity): in 1949 Bergmann launched
the program of canonically quantizing the full nonlinear theory of
general relativity (promoting in particular the full metric
$g_{\mu\nu}$ to a quantum operator, rather than the deviations
from the flat spacetime metric), starting with the development of
a Hamiltonian formulation of constrained systems
\cite{Berg1,Berg2} (for historical studies see
\cite{Salisbury:2012ona,Salisbury:2007br}). The problem of
constrained Hamiltonian dynamics was also pursued by Dirac
\cite{DiracHam} (who was initially motivated by the relation
between Hamiltonian dynamics and \emph{special} relativity
\cite{DiracForms}); his formalism was then applied to general
relativity by Pirani and Alfred Schild \cite{Schild}.

\subsection{Nonlinearity and curved spacetime}

Let us now focus on Feynman's approach to the derivation of
general relativity from the theory of a spin-$2$ graviton field.
The starting point is the linearized theory for such a field;
nonlinearity then arises from the fact that the graviton has to
couple universally with anything carrying energy-momentum,
including itself. General covariance, along with the usual
geometric interpretation of general relativity, comes about merely
as interesting and useful (albeit somewhat mysterious) byproducts,
which can be seen as related to gauge invariance. In Feynman's
words:
\begin{quote}
The fact is that a spin-two field has this geometric
interpretation; this is not something readily explainable -- it is
just marvelous. The geometric interpretation is not really
necessary or essential to physics. It might be that the whole
coincidence might be understood as representing some kind of gauge
invariance (\cite{Feynman:1996kb}, p. 113).
\end{quote}
Besides the understanding of another part of nature in his own and
original way, Feynman sought a fast track toward the quantization
of gravity, which he just regarded as the quantization of another
field. In particular, once general relativity had been understood
as the result of the interactions of spin-$2$ quanta, quantum
gravity effects would be taken into account by including diagrams
with closed loops. By treating gravity in this way, difficult
conceptual and technical issues concerning the meaning of quantum
geometry would not show up (\cite{WheelerFest2}, p. 377).

Feynman's justification for his starting point was the fact that
gravity behaves as a ${1}/{r^2}$ force, through which like charges
(i.e. masses) attract. By trial and error, one would then arrive
at the hypothesis that such a force is mediated by a new massless
spin-$2$ field (see Section \ref{further} for Feynman's later
arguments for this statement). Then the action would
be\footnote{The original equations appearing in \cite{ChapelHill}
are schematic and present several index and sign mistakes;
however, this fact does not affect the reasoning we are reporting;
in the following we write the correct versions of such
equations.}:
\begin{eqnarray}\label{secondOrAction}
\int \left(\frac{\partial A_{\mu}}{\partial x_{\nu}}-\frac{\partial A^{\nu}}{\partial x^{\mu}}\right)^2 \drm^4x + \int A_{\mu}j^{\mu} \drm^4x + \frac{m}{2}\int \dot{z}_{\mu}^2 \drm s + \frac{1}{2}\int T_{\mu\nu}h^{\mu\nu} \drm^4x \nonumber \\
+ \int (\textrm{second power of first derivatives of $h$} ),
\end{eqnarray}
where $h_{\mu\nu}$ is the new field, satisfying second order
equations of the kind:
\begin{eqnarray} \label{fieldEq}
{h^{\mu\nu,\sigma}}_{,\sigma} - 2
{{\overline{h}^{\mu}}_{\sigma}}^{,\nu\sigma}=
\overline{T}^{\mu\nu}.
\end{eqnarray}
Here the bar operation on a general second rank tensor
$X_{\mu\nu}$ is defined as:
\begin{eqnarray} \label{barOperation}
\overline{X}_{\mu\nu}=\frac{1}{2} \left( X_{\mu\nu} + X_{\nu\mu}
\right) - \frac{1}{2} \eta_{\mu\nu} {X^{\sigma}}_{\sigma}.
\end{eqnarray}
The corresponding equation of motion for particles moving in this
field would be:
\begin{eqnarray} \label{particleEq}
g_{\mu\nu} \ddot{z}^{\nu} = -\left[\rho\sigma,\mu
\right]\dot{z}^{\rho}\dot{z}^{\sigma},
\end{eqnarray}
where $g_{\mu\nu}=\eta_{\mu\nu}+h_{\mu\nu}$, and
$\left[\rho\sigma,\mu \right]$ are the Christoffel symbols of the
first kind. Eqs. (\ref{fieldEq}) are nothing but the linearized
Einstein equations. Further pursuing the analogy with
electromagnetism (for which, as said, the same approach can be
adopted), a crucial fact to be noticed is that, in that case, the
Maxwell equations automatically imply that the electric current
$j^{\mu}$ is conserved. Then, also in the gravitational case, a
suitable $T_{\mu\nu}$, such that the condition
$\partial_{\nu}T^{\mu\nu}=0$ is satisfied, must be found. However,
a problem arises if particles move according to
Eq.(\ref{particleEq}) or, what is the same, if the field
$h_{\mu\nu}$ is coupled to the matter, since the corresponding
$T_{\mu\nu}$ does {\it not} obey to a conservation law. Thus, the
linear theory leads to a consistency problem\footnote{In fact, the
stress-energy tensor $T_{\mu\nu}$ has been specified only in terms
of matter. As such it does not include the energy of the
gravitational field itself. This can be taken into account only as
a nonlinear effect (gravity has to couple with itself), which
would be important also in explaining the precession of the
perihelion of Mercury (see the discussion in
\cite{Feynman:1996kb}, p. 75).}. The addition of a further
contribution $t_{\mu\nu}$ to the stress-energy tensor due to the
gravitational field itself, thus replacing the term $\int
T_{\mu\nu}h^{\mu\nu} \drm^4x$ in the action (\ref{secondOrAction})
with $\int \left(T_{\mu\nu} + t_{\mu\nu}\right)h^{\mu\nu}
\drm^4x$, does not solve the problem, since a variation of $h$
would give new terms. According to Feynman (see \cite{ChapelHill},
p. 274 and, for a more detailed derivation \cite{Feynman:1996kb},
pp. 78-79), a working solution could be obtained only by adding to
the action a nonlinear third order term in $h_{\mu\nu}$, which
gives for $T_{\mu\nu}$ the following equation:
\begin{eqnarray} \label{tensorEq}
g_{\mu\lambda} T^{\mu\nu}_{\;\;\;\;\,,\nu} =-
\left[\rho\nu,\lambda \right]T^{\rho\nu}.
\end{eqnarray}
One could then proceed to the next order approximation, and so
on. However, as a matter of fact, finding the general solution of
Eq. (\ref{tensorEq}) is a very difficult task, which may be
pursued by finding an expression that is invariant under the
following infinitesimal transformation of the tensor field
$g_{\mu\nu}$:
\begin{eqnarray} \label{infinitesimal1}
g_{\mu\nu}' = g_{\mu\nu} + g_{\mu\lambda}\frac{\partial
A^{\lambda}}{\partial x^{\nu}} + g_{\nu\lambda}\frac{\partial
A^{\lambda}}{\partial x^{\mu}} + A^{\lambda}\frac{\partial
g_{\mu\nu}}{\partial x^{\lambda}}.
\end{eqnarray}
Here the $4$-vector $A^{\lambda}$ is the generator of the
transformation. This is a geometric transformation on a Riemannian
manifold, hence one might finally argue that geometry gives the
metric. However, as noticed by Feynman,
\begin{quote}
this would be a marvelous suggestion but it would be made at the
end of the work and not at the beginning. What does one gain by
looking at the problem in this manner? Obviously, one loses the
beauty of geometry but this is not primary. What is primary is
that one had a new field and tried his very best to get a spin-two
field as consistent as possible (\cite{ChapelHill}, p. 275).
\end{quote}
Feynman would later develop the above sketched procedure in great
detail, managing in the end to obtain Einstein's full nonlinear
gravitational field equations. He presented the full derivation in
his graduate course on gravitation (\cite{Feynman:1996kb},
lectures 3-6). In that course, after completing the task and
before switching to applications\footnote{These include the
Schwarzschild solution and wormholes (Lecture 11), cosmology
(Lectures 12-13), supermassive stars (Lecture 14) and black holes
(Lecture 15), closing with the already cited Lecture 16 on
gravitational radiation.}, he devoted some lectures to the usual
geometric approach to gravity (\cite{Feynman:1996kb}, Lectures
7-10). He had already shown that his formalism was able to
reproduce physical effects that in the standard picture are
ascribed to curved spacetime geometry. For example
(\cite{Feynman:1996kb}, pp. 66-69) Feynman showed, by studying the
action of a scalar field, that a constant weak gravitational
field, described by the metric component $g_{44}=1+\epsilon$,
$g_{ii}=-1$, $i=1,2,3$, is exactly reproduced by substituting
$t\rightarrow t'=t\sqrt{1+\epsilon}$, i.e. by time dilation. He
also noticed that this effect is pivotal in getting the right
precession of Mercury's perihelion, which he had computed
previously (\cite{Feynman:1996kb}, pp. 63-65) without taking it
into account and hence getting a result which was $4/3$ of the
right one.

Feynman was of course intrigued by the double nature of
gravitation, and tried to link his approach to the usual one:
\begin{quote}
Let us try to discuss what it is that we are learning in finding
out that these various approaches give the same results
(\cite{Feynman:1996kb}, p. 112).
\end{quote}
In particular, he discussed the issue of flatness of space, since
\begin{quote}
The point of view we had before was that space is describable as
the space of Special Relativity [...] there might be gravity
fields $h_{\mu\nu}$ which have the effect that rulers are changed
in length, and clocks go at faster or slower rates. So that in
speaking of the results of experiments we are forced to make
distinctions between the scales of actual measurements, physical
scales, and the scales in which the theory is written [...] It may
be convenient in order to write a theory in the beginning to
assume that measurements are made in a space that is in principle
Galilean, but after we get through predicting real effects, we see
that the Galilean space has no significance.
(\cite{Feynman:1996kb}, p. 112).
\end{quote}
To better explain the situation, he considered the analogy with an
observer making length measurements on a hot plate, since a ruler
can be affected by temperature. However, in this situation the
observer can resort to a different instrument, such as light,
which is not affected by temperature. On the other hand, gravity
is universal, so
\begin{quote}
we know of no scale that would be unaffected--there is no
``light'' unaffected by gravity with which we might define a
Galilean coordinate system. Thus, all coordinate systems are
equivalent, and they differ only in that different values for the
fields are necessary for the description of clock rates or length
scales (\cite{Feynman:1996kb}, p. 112-113).
\end{quote}
In Section 8.3 of \cite{Feynman:1996kb}, Feynman explored possible
directions to understand how gravity can be both geometry and a
field, and argued (see the quote at the beginning of this Section)
that such a link is provided by gauge invariance. In order to
connect gravity and gauge invariance, one may look for a procedure
to obtain the invariance of the equations of physics under
spacetime dependent coordinate displacements. This amounts to
adding to the Lagrangian new terms involving a gravitational
field. The latter thus emerges as the gauge field enforcing
invariance with respect to local displacements.

\subsection{Relation of Feynman's approach to analogous work}

Unlike previous partial attempts which addressed only the
linearized quantum theory \cite{Fierz:1939ix,Ros1,Ros2,Bronstein}
or iterative arguments, able in principle to generate infinite
nonlinear terms in the Lagrangian as well as in the stress-energy
tensor, but still incomplete \cite{Gupta:1954zz}, Feynman
succeeded in obtaining the full nonlinear Einstein equations by
means of a consistency argument. According to Preskill and Kip S.
Thorne \cite{LectGravPreface}, it is likely that Feynman was
completely unaware of Kraichnan's work as well as Gupta's, thus he
devised this method independently, besides getting more complete
results.

Other people who pursued analogous approaches to gravity include
Steven Weinberg, whose approach was quite different from Feynman's
one, being based on analiticity properties of graviton-graviton
scattering amplitudes (Weinberg also derived Maxwell's equations
along the same lines) \cite{Weinberg:1964kqu,Weinberg:1965rz}, and
Deser \cite{Deser:1969wk}-\cite{Deser:2009fq}. Unlike Weinberg's,
Deser's approach was closely analogous to Feynman's one, but more
elegant, and it was in fact the first published completion of the
program started with Kraichnan and Gupta. Moreover, Deser applied
his method also to Yang-Mills theory. The relation of Lorentzian
spin-2 field theories with general covariance was later
investigated in a more rigorous and general way by Robert M. Wald
\cite{Wald:1986bj}.

\subsection{A neutrino-induced gravity?}\label{Neutrino}

Feynman concluded his comments at Chapel Hill with some
speculations on a possible theory of gravitation built on already
known fields, i.e. not relying on a brand new spin-2 field, a
promising candidate being the neutrino, described by a weakly
coupled field with (at that time) zero rest mass. But how could a
gravitational-like force be obtained by exchange of neutrinos? The
exchange of a single neutrino is ruled out by its half-integer
spin, which requires orthogonal initial and final states;
similarly, exchange of two neutrinos has to be discarded, because
the resulting potential would fall off faster than ${1}/{r}$. A
possible solution emerges from exchanging one neutrino between two
bodies while, in turn, each body exchanges one neutrino with the
rest of the Universe, located at a fixed distance. This situation,
however, gives rise to a logarithmic divergence in the potential,
and a higher order divergence would be obtained if one considered,
for instance, four neutrino processes. Thus, Feynman was led to
the conclusion that a theory built of neutrinos would not be
viable: ``This is obviously no serious theory and is not to be
believed" (\cite{ChapelHill}, p. 276). The inadequacy of a theory
of gravitation involving neutrinos would be later discussed in
much more detail in the Lectures on Gravitation
(\cite{Feynman:1996kb}, Lecture 2, pp. 23-28).

\subsection{Further work}\label{further}

After 1957, Feynman's investigations concentrated on the quantum
theory of gravity, addressing in particular the issues of loop
diagrams, renormalization and unitarity, and also on the
applications of relativistic gravity to astrophysics and
cosmology\footnote{Such applications, indeed, had meanwhile begun
to flourish, with the discovery of a wealth of phenomena, such as
the cosmic microwave background, pulsars and quasars (see e.g.
\cite{Thorne:1994xa} and references therein).}. These applications
are described, as we saw, in the Caltech lectures
\cite{Feynman:1996kb}, and further discussed in the Hughes
lectures \cite{FeynmanHughes1}. Also in this case, Feynman
obtained some original results. However, we do not address
applicative topics in this paper, leaving them to a future
publication.

Concerning quantum gravity, Feynman gave a preliminary assessment
of his progress at the La Jolla conference\footnote{The
International Conference on the Theory of Weak and Strong
Interactions was held in June 14-16 1961 at the University of
California, San Diego, in La Jolla. We recall that, here, Geoffrey
F. Chew gave his celebrated talk on the $S$-matrix \cite{Chew},
while an afternoon session was devoted to the theory of
gravitation, with Feynman reporting on his work on the
renormalization of the gravitational field and recognizing
non-unitarity as the main difficulty, shared also by Yang-Mills
theory.} in 1961 \cite{LaJolla} and in the already mentioned
letter to Weisskopf \cite{WeisskopfLetter} where, consistently
with his views, classical gravitational radiation is mainly
discussed in a quantum field theoretical framework, but within the
tree approximation, noting that neglecting radiative corrections,
the related problems with divergencies and unitarity are not
present or, in his words (\cite{WeisskopfLetter}, p.1), ``without
the radiation correction there is no difficulty''. A fuller
account was given in the famous 1963 Warsaw talk (see next
Section). Interestingly enough, the written version of that talk
was the first paper that he published on gravity, despite an
interest which at the time had already been going on for a decade.
After that, Feynman only published two more papers on the subject
\cite{WheelerFest1,WheelerFest2}, in the Festschrift for Wheeler's
60th birthday \cite{Klauder:1972je}, in which he completed the
discussion of his Warsaw paper, giving many details. Apparently,
the Wheeler festschrift papers present research performed several
years before their publication, which Feynman had been reluctant
to publish \cite{LectGravPreface} due to its presumed
incompleteness\footnote{Also, starting from the late 1960s,
Feynman became more and more absorbed in studying partons and
strong interactions, which were his main interest in the 1970s.}.
For the same reason, he did not authorize the distribution of the
notes for the last 11 lectures, addressing the quantization of
gravity, which are therefore not included in Ref.
\cite{Feynman:1996kb}. As already remarked, the 1966-67 Hughes
lectures on astronomy, astrophysics and cosmology
\cite{FeynmanHughes1} include a presentation of general
relativity. Notably, an elementary discussion of this topic is
given also in the undergraduate Caltech lectures
\cite{Feynman:1963uxa}.

Both in the Caltech and in the Hughes lectures, Feynman gave some
further justification for the choice of a spin-$2$ field, based on
the following observation. Unlike the electric charge, which is
the source of the electromagnetic force, the source of the
gravitational force -- energy-momentum -- is not a relativistic
invariant, but rather grows with the velocity. Feynman noted that
the charge associated with a spin-$0$ field would instead decrease
with the velocity. This result can be traced back to an old
argument by Einstein (which he never published, but was recalled
by him in \cite{Einstein1933}), according to which, in a
Lorentz-covariant scalar theory of gravity (which is what must be
expected to descend from a Lorentzian quantum field theory of
spin-$0$ particles), the vertical acceleration of a body would
depend on its horizontal velocity (and also on its internal and/or
rotational energy)\footnote{This was the reason for Einstein's
initial rejection of Lorentz-invariant theories of gravity based
on a scalar field, and for his criticism toward the first theory
proposed by Gunnar Nordstr\"{o}m \cite{Nordstrom} (see
\cite{Norton} for details).}. Moreover, a scalar theory of gravity
cannot account for the observed phenomenon of light deflection.
Hence the possibility of a spin-0 field is ruled out. Feynman thus
singled out the spin-2 case.

\section{Quantum corrections}

The issue of loop corrections and renormalization in quantum
gravity is publicly addressed by Feynman for the first time (to
the best of our knowledge) in 1961, at the already mentioned
Conference in La Jolla \cite{LaJolla}. Here he recognized
nonlinearity as a source of difficulty for both gravitation and
Yang-Mills theory. Basically, the sources of the gravitational
field are energy and momentum, which are locally conserved, and
the gravitational field carries energy and momentum itself, so it
is self-coupled. Similarly, the source of a Yang-Mills field is
the isotopic spin current, which is locally conserved, and the
Yang-Mills field carries isotopic spin itself, and thus it is also
self-coupled. In both cases, the result is a nonlinear field
theory.

\subsection{Attacking the problem}

As anticipated, all of Feynman's results on quantum gravity are
essentially contained in the already mentioned Ref.
\cite{Feynman:1963ax}, a report of the talk given at the
International Conference General Relativity and Gravitation (GR3),
held in Warsaw in 1962\footnote{Interestingly, some notes that
apparently refer to the Warsaw talk are included in Feynman's
notes for the famous Caltech lectures, which were given in the
same period. In particular, in Folder 40.5, which is available
online \cite{Notes}, the last two pages specifically discuss
gravity, Yang-Mills theory and loop corrections, with the heading
``Grav talk''. This further shows how deeply involved in these
matters he was in that period.}$^{,}$\footnote{As Trautman
recalled in some recently published memories (\cite{Trautman}, p.
406), the text of Feynman's plenary lecture at the GR3 conference
became available too late to be included in the proceedings,
therefore it was published as a regular article in Acta Physica
Polonica in 1963.}, and in the two Wheeler festschrift papers
\cite{WheelerFest1,WheelerFest2}. According to his original
strategy, in \cite{Feynman:1963ax} Feynman did not dwell on the
problem of the quantization of space-time geometry, but rather
constructed a quantum field theory for a massless spin-2 field --
the graviton -- and then worked out the results at different
perturbative orders, with quantum corrections being taken into
account by including loop diagrams. Since the focus was on the
quantum theory, the Einstein equations and the corresponding
action were assumed from the beginning, rather than derived as in
the Lectures \cite{Feynman:1996kb}, and then quantized by adopting
a standard procedure.

In the introduction of \cite{Feynman:1963ax}, Feynman outlined his
aims and approach once again:
\begin{quote}
My subject is the quantum theory of gravitation. My interest in it
is primarily in the relation of one part of nature to another.
There's a certain irrationality to any work in gravitation, so
it's hard to explain why you do any of it; for example, as far as
quantum effects are concerned let us consider the effect of the
gravitational attraction between an electron and a proton in a
hydrogen atom; it changes the energy a little bit. Changing the
energy of a quantum system means that the phase of the wave
function is slowly shifted relative to what it would have been
were no perturbation present. The effect of gravitation on the
hydrogen atom is to shift the phase by 43 seconds of phase in
every hundred times the lifetime of the Universe! [...] I am
limiting myself to not discussing the questions of quantum
geometry nor what happens when the fields are of very short wave
length. [...] I suppose that no wave lengths are shorter than
one-millionth of the Compton wave length of a proton, and
therefore it is legitimate to analyze everything in perturbation
approximation; and I will carry out the perturbation approximation
as far as I can in every direction, so that we can have as many
terms as we want (\cite{Feynman:1963ax}, p. 697).
\end{quote}
Again, Feynman advocated a substantial unity of nature, which
required to reconcile gravity and quantum mechanics, although he
admitted that the work had some irrationality in it due to the
smallness of the effects. Consistently with what he said at Chapel
Hill concerning the choice between mathematical rigor and thought
experiments, he chose the latter, pursuing a perturbative
approach, solving simple specific problems and later switching to
more complex ones:
\begin{quote}
So please appreciate that the plan of the attack is a
succession of increasingly complex physical problems; if I could
do one, then I was finished, and I went to a harder one imagining
the experimenters were getting into more and more complicated
situations (\cite{Feynman:1963ax}, p. 698).
\end{quote}

\subsection{A perturbative approach}

As said, the starting point is the Einstein-Hilbert Lagrangian for
gravity coupled to scalar matter:
\begin{quote}
I started with the Lagrangian of Einstein for the interacting
field of gravity and I had to make some definition for the matter
since I'm dealing with real bodies and make up my mind what the
matter was made of; and then later I would check whether the
results that I have depend on the specific choice or they are more
powerful (\cite{Feynman:1963ax}, p. 698). I can do only one
example at a time. I took spin zero matter.
\end{quote}
The metric is split in the following way:
\begin{eqnarray}\label{metric}
g_{\mu\nu} = \delta_{\mu\nu} + \kappa h_{\mu\nu},
\end{eqnarray}
which allows, upon substitution and subsequent expansion, to cast
the Lagrangian in the form:
\begin{eqnarray} \label{lagrangian}
L &=& \int  \left(h_{\mu\nu,\sigma} \overline{h}_{\mu\nu,\sigma} - 2 \overline{h}_{\mu\sigma,\sigma}\overline{h}_{\mu\sigma,\sigma}\right) \drm \tau + \frac{1}{2} \int \left( \phi_{,\mu}^2 - m^2 \phi^2 \right) \drm \tau \nonumber \\
& & + \kappa \int
\left(\overline{h}_{\mu\nu}\phi_{,\mu}\phi_{,\nu} - m^2
\frac{1}{2}{h}_{\sigma\sigma}\phi^2 \right) \drm \tau + \kappa
\int \left( hhh \right)\drm \tau + \kappa^2 \int \left( hh\phi\phi
\right) \drm \tau + ...
\end{eqnarray}
Here again the bar operation (\ref{barOperation}) is used, and a
schematic notation is adopted for the highly complex higher order
terms. The first two terms are the free Lagrangians of the
gravitational field and of matter, respectively. The first step is
to solve the problem classically, which is carried out by varying
the Lagrangian (\ref{lagrangian}) with respect to $h$ and, then,
to $\phi$. The following equations of motion with a source term
are obtained:
\begin{eqnarray}
h_{\mu\nu,\sigma\sigma} - \overline{h}_{\sigma\nu,\sigma\mu} -
\overline{h}_{\sigma\mu,\sigma\nu} &=& \overline{S}_{\mu\nu}
\left(h, \phi \right),\label{source1}\\
\phi_{,\sigma\sigma} - m^2 \phi &=& \chi \left(\phi, h \right).
\label{source2}
\end{eqnarray}
The next step consists on obtaining the propagators, by following
a procedure analogous to that used in electromagnetism. But soon
Feynman realized that Eq. (\ref{source1}) is singular, so he
resorted to the invariance of the Lagrangian under the
transformation:
\begin{eqnarray} \label{invariance}
h_{\mu\nu}^{'} = h_{\mu\nu} + 2 \xi_{\mu,\nu} + 2 h_{\mu\sigma}\xi_{\sigma,\nu} + \xi_{\sigma} h_{\mu\nu,\sigma},
\end{eqnarray}
where $\xi_{\mu}$ is arbitrary. As a consequence, the consistency
of Eq. (\ref{source1}) requires the source $S_{\mu\nu}$ to have
zero divergence, since the symmetry (\ref{invariance}) implies the
identical vanishing of the divergence of the barred left hand side
of Eq. (\ref{source1}). By making the simple gauge choice
$\overline{h}_{\mu\sigma,\sigma}=0$, finally the law describing
the gravitational interaction of two systems by means of the
exchange of a virtual graviton can be obtained. In order to
achieve a higher accuracy, one has then to work out radiative
corrections. Besides working out the propagator\footnote{In
Feynman's approach to quantum field theory propagators are just
Green's functions for the classical field equations, with suitable
boundary conditions.}, Feynman gave other examples of calculations
with diagrams, explicitly computing the amplitude for the coupling
of two particles to a graviton, namely an interaction vertex, and
finally considering the gravitational analogue of the Compton
effect, where the photon is replaced by a graviton.

\subsection{One-loop corrections}

Feynman then switched to more complex situations, which required
to go beyond the tree-diagram approximation:
\begin{quote}
However the next step is to take situations in which we have what
we call closed loops, or rings, or circuits, in which not all
momenta of the problem are defined (\cite{Feynman:1963ax}, pp.
703-704).
\end{quote}
Unfortunately, the inclusion of loops brings in several new
conceptual issues, as clearly expressed by Feynman's words:
\begin{quote}
This made me investigate the entire subject in great detail to
find out what the trouble is. I discovered in the process two
things. First, I discovered a number of theorems, which as far as
I know are new, which relate closed loop diagrams and diagrams
without closed loop diagrams (\emph{sic}) (I shall call the latter
diagrams ``trees''). The unitarity relation which I have just been
describing, is one connection between a closed loop diagram and a
tree; but I found a whole lot of other ones, and this gives me
more tests on my machinery. So let me just tell you a little bit
about this theorem, which gives other rules. It is rather
interesting. As a matter of fact, I proved that if you have a
diagram with rings in it there are enough theorems altogether, so
that you can express any diagram with circuits completely in terms
of diagrams with trees and with all momenta for tree diagrams in
physically attainable regions and on the mass shell. The
demonstration is remarkably easy (\cite{Feynman:1963ax}, p. 705).
\end{quote}
In the discussion section of \cite{Feynman:1963ax} (pp. 714-717),
in answering DeWitt's questions about the statement of the tree
theorem and the nature of the proof for the one-loop case, Feynman
gave more details about his arguments. The first seed of his idea
was related to the computation of the self-energy in quantum
electrodynamics and, in particular, of the Lamb shift of the
hydrogen atom, but he really worked out the whole machinery when
dealing with the quantum theory of gravitation.

By working out in detail one-loop calculations, Feynman soon
realized that unitarity was lost because some contributions
arising from the unphysical longitudinal polarization states of
the graviton did not cancel. As suggested by
Gell-Mann\footnote{This circumstance is also mentioned in Ref.
\cite{Gell-Mann}.}, he considered a similar problem in the simpler
context of Yang-Mills theory, and found the same pathologic
behavior:
\begin{quote}
But this disease which I discovered here is a disease which exist
in other theories. So at least there is one good thing: gravity
isn't alone in this difficulty. This observation that Yang-Mills
was also in trouble was of very great advantage to me. [...] the
Yang-Mills theory is enormously easier to compute with than the
gravity theory, and therefore I continued most of my
investigations on the Yang-Mills theory, with the idea, if I ever
cure that one, I'll turn around and cure the other
(\cite{Feynman:1963ax}, p. 707).
\end{quote}
In order to solve this tricky issue, Feynman succeeded in showing
that any diagram with closed loops can be expressed in terms of
sums of on shell tree diagrams, notwithstanding the fact that the
process of opening a loop by cutting a graviton line implies
replacing a virtual graviton with a real transverse one. This is
the very content of his tree theorem (which was fully treated in
\cite{WheelerFest1}). In order to guarantee gauge invariance, one
has to sum the whole set of tree diagrams corresponding to a given
process. Feynman then hinted to the general problem, and commented
that his novel procedure works fine for the one-loop case:
\begin{quote}
The question is: Can we express the closed ring diagrams for some
process into a sum over various other processes of tree diagrams
for these processes? Well, in the case with one ring only, I am
sure it can be done, I proved it can be done and I have done it
and it's all fine. And therefore the problem with one ring is
fundamentally solved; because we say, you express it in terms of
open parts, you find the processes that they correspond to,
compute each process and add them together (\cite{Feynman:1963ax},
p. 709).
\end{quote}
It is of interest to notice that Feynman's tree theorem has
recently been revived in the context of advanced perturbative
calculations and generalized unitarity (see e.g.
\cite{Britto:2010xq}-\cite{Maniatis:2016nmc}).

Finally, Feynman asked himself how to get the same result by
integrating the closed loop directly, promptly giving the answer.
On the one hand, a further term (like a mass term) has to be added
to the Lagrangian in order to make it non-singular, but such a
term breaks gauge invariance; on the other hand a contribution,
obtained by taking a ghost particle (with spin-$1$ and Fermi
statistics) going around the ring and artificially coupled to the
external field, has to be subtracted. In this way unitarity and
gauge invariance would be recovered. The same calculation can be
carried out for the Yang-Mills theory, but in this case the ghost
particle must have spin-$0$ (all the results on Yang-Mills theory
would be written up in Ref. \cite{WheelerFest2}). This discovery
would have been of seminal importance for gauge theories. Indeed,
Feynman's formalism was later fully developed by DeWitt who,
inspired by Feynman's talk in Warsaw and by the discussion that
followed, solved the problem of extending it to two
\cite{DeWitt:1964yg} and arbitrarily many \cite{DeWitt:1967ub,
DeWitt:1967uc} loops, while Ludvig D. Faddeev and Victor N. Popov
\cite{FaddeevPopov} derived the extra terms to be added to the
Lagrangian in a much simpler way, by means of a functional
integral quantization rather than Feynman diagrams.

\subsection{Renormalizability}

Feynman's conclusive remarks about quantum corrections deal with
the possibility of extending the above results to higher orders
(e.g. two or more loops):
\begin{quote}
Now, the next question is, what happens when there are two or more
loops? Since I only got this completely straightened out a week
before I came here, I haven't had time to investigate the case of
2 or more loops to my own satisfaction. The preliminary
investigations that I have made do not indicate that it's going to
be possible so easily gather the things into the right barrels.
It's surprising, I can't understand it; when you gather the trees
into processes, there seems to be some loose trees, extra trees
(\cite{Feynman:1963ax}, p. 710).
\end{quote}
Clearly, he did not know how to manage the problem. The same
feeling was expressed in the Lectures on Gravitation:
\begin{quote}
I do not know whether it will be possible to develop a cure for
treating the multi-ring diagrams. I suspect not -- in other words,
I suspect that the theory is not renormalizable. Whether it is a
truly significant objection to a theory, to say that it is not
renormalizable, I don't know (\cite{Feynman:1996kb}, Lecture 16,
pp. 211-212).
\end{quote}
In the same lecture, Feynman pointed out that the Yang-Mills case
constituted a simpler context in which to tackle the same
difficulties, whose source is mainly the lack of unitarity of some
sums of diagrams. Then he argued for a substantial
non-renormalizability of the theory, probably as a consequence of
these difficulties\footnote{We point out that, to the best of our
knowledge, it is not clear if here Feynman hinted to a link
between non-unitarity and non-renormalizability issues. We now
know that there is no such link, since while unitarity at
arbitrarily many loops would have been shown later by DeWitt
\cite{DeWitt:1967ub, DeWitt:1967uc}, in the case of gravity this
does not imply renormalizability. Conversely, modified theories of
gravity whose action is quadratic in the curvature have been shown
to be renormalizable, but not unitary \cite{Stelle:1976gc} (see
however \cite{Anselmi:2017ygm} for a proposal of a quantum theory
of gravity which is both unitary and renormalizable).}. In fact,
Yang-Mills theory was later shown to be renormalizable
\cite{tHooft:1971akt}-\cite{tHooft:1972qbu}, while the divergences
of gravity proved to be too strong to be tamed by renormalization
\cite{tHooft:1974toh}-\cite{vandeVen:1991gw}, confirming Feynman's
suspect in that case. It is noteworthy that, against the common
lore of the time, Feynman was not convinced that
non-renormalizability meant that a theory was
inconsistent\footnote{This is also recalled by Gell-Mann in Ref.
\cite{Gell-Mann}, where he states that ``he was always very
suspicious of unrenormalizability as a criterion for rejecting
theories''.}. As he himself declared in one of his last interviews,
given in January 1988 (quoted in \cite{Mehra:1994dz}, p. 507):
\begin{quote}
The fact that the theory has infinities never bothered me quite so
much as it bother others, because I always thought that it just
meant that we've gone too far: that when we go to very short
distances the world is very different; geometry, or whatever it
is, is different, and it's all very subtle.
\end{quote}
Modern views on quantum field theory, which emerged in the 1970s,
indeed see non-renormalizability to be not a significant objection
against a theory, but rather as a signal that the theory loses
validity at energies higher than a certain scale, i.e. it is an
\emph{effective field theory} (see for example
\cite{Weinberg:2021exr,Weinberg:2016kyd} for historical
discussions), which can however be used to make predictions below
that scale. As it is now well understood, this is just the case of
gravity (see e.g. \cite{Burgess:2003jk} and references therein).
However, Feynman apparently never embraced this view. In
\cite{PreskillTalk} (slide 37), Preskill remembers:
\begin{quote}
I spoke to Feynman a number of times about renormalization theory
during the mid-80s (I arrived at Caltech in 1981 and he died in
1988). I was surprised on a few occasions how the effective field
theory viewpoint did not come naturally to him''. [...] Feynman
briefly discusses in his lectures on gravitation (1962) why there
are no higher derivative terms in the Einstein action, saying this
is the ``simplest'' theory, not mentioning that higher derivative
terms would be suppressed by more powers of the Planck length.
\end{quote}

\section{The Hughes lectures}

The 1966-67 Hughes lectures\footnote{Like other sets of lectures
given in those years, from 1966 to 1971, the presently analyzed
ones have been made available to us by John T. Neer \cite{Hughes},
who rewrote the notes he had taken attending the lectures and also
undertook the task of including some up to date information when
needed. These notes were not revised by Feynman, as Neer admits
\cite{Hughes}: ``these are my personal notes and are only a
representation of the lectures I attended. They are to the best of
my ability my recreation from memory and my original real time
notes. No AV recording system was used in the transcription of my
raw notes''. } on Astronomy and Astrophysics \cite{FeynmanHughes1}
are of great interest not only as a further example of Feynman's
curiosity in action in a different field, but also because they
contribute to our understanding of Feynman's ideas about gravity.
Indeed, an important part of them is devoted to introducing the
elements of general relativity, with the aim of applying it to
astrophysics and cosmology. In the following we shall therefore
focus on the third chapter, which is the relevant
one\footnote{Notice that the chapter numeration in the table of
contents at the beginning of the lectures does not match the
numeration in the main text. Here we refer to the latter.}. The
remaining chapters concern the internal structure of stars, their
evolution, active galactic nuclei, and the solar system.

The lectures begin with an overview of the subject matter,
organized by growing scale, i.e. from the solar system to quasars.
After that, Feynman deals with the Universe as a whole, namely
cosmology. This is the subject of the last part of Chapter 1 and
of Chapter 2 of the notes, in which a model Universe treated using
Newtonian gravity is discussed. Then, in Chapter 3, Feynman
started discussing general relativity, by criticizing the
geometric approach (cf. the quotation we reported in Section 2.1)
and giving some motivation about why gravity should be described
by a spin-2 field, along the lines of the Caltech lectures
\cite{Feynman:1996kb} (\cite{FeynmanHughes1}, pp. 30-32). After
that he switched to the usual, Einsteinian views. His choice was
probably motivated by the fact that the field theoretical
derivation would be both too advanced for the audience, and too
long for a course whose main focus was on applications. For each
step he properly credited Einstein, but, interestingly, he often
gave a quite original motivation, different from Einstein's one
(this is especially intriguing for the case of light deflection,
as we discuss below). He explained the different notions of mass,
the equivalence principle (stated by the usual elevator
thought-experiment), and gravitational redshift
(\cite{FeynmanHughes1}, pp. 32-35). Then (\cite{FeynmanHughes1},
p. 36) he included an interesting comment on the well-known
problem of the consistency of the principle of equivalence with
the fact that accelerated electric charges radiate\footnote{This
topic is the source of apparent paradoxes (see e.g.
\cite{RohrlichBook} for a review). For example, a charge sitting
at rest in a gravitational field is not seen to radiate, despite
being indistinguishable from a uniformly accelerated charge
according to the equivalence principle, thus allowing an observer
at rest to distinguish gravitational pull from other kinds of
acceleration.}. This problem had been solved a few years before
Feynman's lectures, by theorists like Fritz Rohrlich
\cite{FultonRohrlich,Rohrlich} and DeWitt
\cite{DeWitt:1960fc,DeWitt:1964de}, who showed that the
equivalence principle is not violated by realizing that the usual
laws of electrodynamics hold in inertial frames, while when
transformations between accelerating frames are involved, whether
a charge radiates or not is actually an observer-dependent
statement\footnote{It was later shown \cite{Boulware,deAlmeida}
that the radiation from a uniformly accelerated charge is entirely
emitted beyond the Rindler horizon of a co-accelerating observer,
which therefore cannot observe it.}. Feynman's aim was to use
these results to argue on physical grounds in favor of the
modification of electromagnetism in a gravitational field. This
had been discussed more explicitly in the Lectures on Gravitation
\cite{Feynman:1996kb}, pp. 123-4, where Feynman declared:
``Clearly, some interaction between gravity and electrodynamics
must be included in a better statement of the laws of electricity,
to make them consistent with the principle of equivalence''.
Indeed, such a phenomenon is peculiar to electrodynamics in the
presence of a gravitational field (another one, he noticed, are
the curved trajectories of light rays). In fact, Feynman was just
motivating the standard modification of the inhomogeneous
Maxwell's equations on curved spacetime, which is induced by the
substitution of the ordinary derivatives with covariant ones (this
is the \emph{minimal coupling} with the gravitational field), but
not in the purely geometric and formal way which is commonly found
in textbooks.

In a very interesting discussion (\cite{FeynmanHughes1}, pp.
38-39), Feynman used a variational approach to show that if a
traveling clock maximizes the time elapsed, then it satisfies the
usual Newtonian equation of motion of a particle subject to the
gravitational acceleration $\mathbf{g}$, i.e. if it is freely
falling. This is of course a ``flat space'' version of the
geodesic principle. The same discussion is present in the Caltech
lectures (\cite{Feynman:1996kb}, pp. 94-95), but without any
calculation. This simple computation provided a motivation for the
relativity of time in a gravitational field. Then Feynman argued
that if time intervals are relative, then ``because of the
space-time relationship''(\cite{FeynmanHughes1}, p. 40), space
distances must be relative as well, hence he introduced position
dependent coefficients in the expression of the usual proper time
interval. This further required a ``replacement of Newton's simple
$\mathbf{F}=m\mathbf{a}$ law'' (\cite{FeynmanHughes1}, p. 40),
which is the geodesic equation (this equation is not written down
in its general form, though). These steps had been given in more
detail in the Lectures \cite{Feynman:1996kb} (pp. 137-139). This
discussion was followed by the interpretation of the position
dependent coefficients as the components of the metric tensor of
curved spacetime\footnote{``So we have a real mess on our hand
with 10 potentials all coupled together. So we must try to put
some order to them. [...] We need some geometrical
interpretation'' (\cite{FeynmanHughes1}, p. 41).}, a lengthy
discussion of curved two-dimensional geometry
(\cite{FeynmanHughes1}, pp. 41-44), and a very brief sketch of the
Riemann tensor (without giving its full expression) and of
Einstein's equations as a generalization of the Poisson equation
(\cite{FeynmanHughes1}, p. 52). The latter material was presented
in considerable less detail than in the Lectures
\cite{Feynman:1996kb}. After that, Neer included some additional
material on spacetime, curved metrics and the relation with
accelerated frames taken from Chapter 2 of Weber's book
\cite{WeberBook} (\cite{FeynmanHughes1}, p. 46-47). This book had
not been recommended by Feynman since, as Neer writes in the
preface of \cite{FeynmanHughes1}, he ``never called out a reading
reference''.

The next topic is a description of the Schwarzschild solution of
the field equations (\cite{FeynmanHughes1}, pp. 47-49). There is
no full derivation of it, since Feynman had not developed all the
necessary tools in the previous lectures, therefore Neer refers
again to Weber's book \cite{WeberBook}, (pp. 56-60), for the
missing steps. After a computation of gravitational redshift in
the Schwarzschild metric (\cite{FeynmanHughes1}, pp. 50-51), there
was a lengthier description of the maximum proper time principle
to find trajectories in a gravitational field, which is then
written down explicitly for the Schwarzschild metric case and
applied first to radial motion of light (\cite{FeynmanHughes1}, p.
52), and then to a quite detailed discussion of the orbits
(\cite{FeynmanHughes1}, pp. 53-54b). By contrast, in the Caltech
lectures, such a computation had been performed in the
perturbative approach (\cite{Feynman:1996kb}, pp. 59-61), and
repeated for the full Schwarzschild metric only for radial
geodesics (\cite{Feynman:1996kb}, p. 201). Then, Feynman switched
to light bending by the Sun. However, instead of adapting the
previous discussion to unbound orbits of a massless particle, he
computes the deflection of a particle in Newtonian gravity, and
then puts $v=c$ (\cite{Feynman:1996kb}, p. 54c)\footnote{This
discussion was probably triggered by a question from the audience,
since Feynman says: ``The question arose how the classical bending
of light was calculated and later shown by Einstein to be twice
what it should be''.}. The result here obtained, which is:
\begin{eqnarray}\label{deflection}
\theta=\frac{2GM_S}{ac^2}
\end{eqnarray}
where $M_S$ is the solar mass and $a$ the impact parameter (which
is the radius of the Sun for a grazing light ray), is thus half of
the correct one. Feynman then claimed that the correct result had
been obtained by Einstein by assimilating space with a varying
gravitational potential to a medium with a varying index of
refraction. This statement is actually wrong. While it is true
that Einstein, back in 1911, discussed light deflection in this
way \cite{Einstein1911}, he actually obtained a result coinciding
with the Newtonian one (\ref{deflection}). In fact, at that time
he still had to realize that space was curved, and it is just
space curvature which gives the ``other half'' of the result.
Later, in 1916 \cite{Einstein1916}, Einstein got the correct
result by studying null geodesics. Quite interestingly, Feynman
asserted that Einstein got the following result for a finite mass
particle:
\begin{eqnarray}\label{deflection2}
\theta=\frac{2GM_S}{av^2}\left(1+\frac{v^2}{c^2}\right)
\end{eqnarray}
which of course for $v=c$ gives the correct deflection angle. The
reference to Einstein is not correct, since as far as we know Eq.
(\ref{deflection2}) does not appear in any of Einstein's writings.
Instead, this statement is quite likely connected to a discussion
contained in the Caltech lectures (\cite{Feynman:1996kb}, p. 41),
where Feynman heuristically explained the factor of $2$ as due to
magnetic-like gravitational interactions, which become equally
important as the static ones for an object moving with the speed
of light.

The last topic is a qualitative discussion of the Schwarzschild
radius and of black holes (which are called ``black stars'', as
appropriate for those times) (\cite{FeynmanHughes1}, pp. 55-58).

\subsection{An unpublished approach to the gravitational interaction}

The Hughes lectures on Astronomy and Astrophysics
\cite{FeynmanHughes1} contain some statements that may hint to the
fact that Feynman had been thinking at gravity in a way analogous
to his treatment of electromagnetism, recently uncovered in
\cite{FeynmanHughes2} and \cite{Gottlieb} and discussed in detail
in \cite{DeLuca:2019ija} and \cite{DiMauro:2020bpd}. Indeed,
Feynman developed a formulation of electromagnetism that was
relativistic from scratch, and again having its roots in his
belief that nature is fundamentally quantum, with classical
fundamental interactions emerging from quantum theories. His
treatment starts with the statement that, for electromagnetism,
the relevant charge is unchanged by motion; prominence is given to
the electromagnetic potentials, since they (and not the fields)
are the basic objects in the quantum theory. We have already seen
(in Section 5.5) how a similar reasoning led Feynman to the
conclusion that the relevant potential for gravity had to be a
tensor. At variance with electromagnetism, general relativity is
typically formulated by starting from the potentials (i.e. the
metric), while the analogs of the electric and magnetic fields
(the connection coefficients) are considered as derived
objects\footnote{We should mention, however, that, since the
1970s, a few ``pure connection'' formulations of general
relativity, which are the basis of modern approaches to quantum
gravity, have been developed (see e.g. \cite{Krasnov} and
references therein).}. Thus, it makes sense to develop an approach
to gravity analogous to Feynman's approach to electromagnetism.

Let us briefly describe Feynman's approach to electromagnetism,
before turning to gravity. As described in
\cite{DeLuca:2019ija,Gottlieb}, Feynman used relativity to argue
that the force felt by an electric charge $q$ in the presence of
other charges has to be linear in the velocity, that is, of the
form:
\begin{eqnarray}\label{forcelin}
F_i(\mathbf{x},\mathbf{v})=q(E_i+v_j\,B_{ij}), \qquad i,j =1,2,3,
\end{eqnarray}
where $E_i$ and $B_{ij}$ are coefficients depending on the other
charges. Then, relativistic covariance allowed him to restrict the
force (\ref{forcelin}) to assume the form of the Lorentz force,
and to derive the transformation properties of the coefficients,
which are identified with the usual electric and magnetic fields
(in particular, the magnetic field is obtained as
$B_i=\frac{1}{2}\epsilon_{ijk}B_{jk}$). For gravity, Feynman
argued that the force acting on a massive particle has to be
quadratic in the velocity, rather than linear, due to the higher
tensor character of gravitation with respect to electromagnetism.
Thus the general expression is (\cite{FeynmanHughes1}, p. 33):
\begin{eqnarray}\label{Force}
F_i=m_0(C_i + v_j\beta_{ij}+v_jv_k\delta_{ijk}),\qquad
i,j,k=1,2,3,
\end{eqnarray}
where $C_i$ is the usual Newtonian field, and $m_0$ is the
(gravitational) mass of the given particle. Presumably, this
discussion should motivate -- in Feynman's aims -- why the
geodesic equation is quadratic in the velocity of the moving
particle, as he states:
\begin{quote}
What Einstein did was then to set out to find the laws of
motion and the laws determining the coefficients.
\end{quote}
For electromagnetism, after deriving the Lorentz force, Feynman
switched to Maxwell's equations. The whole derivation is developed
in the 1967-68 Hughes lectures on electromagnetism
\cite{FeynmanHughes2}. In particular, the homogeneous Maxwell
equations are obtained from the observation that an invariant
action for a relativistic particle in a field can be written by
adding to the free action an invariant 4-potential term
(\cite{FeynmanHughes2}, p. 42)\footnote{He also briefly entertains
the possibility of a spacetime scalar potential, which is however
immediately discarded in view of the fact that ``there are no
known laws which are derivable from this action'' (cf. the
discussion in Section 5.4).}:
\begin{eqnarray}\label{RelAct2}
S = \int_{\tau_i}^{\tau_f} (-m_0 c^2 \, {\rm d}\tau - q A_{\mu} \,
{\rm d}x^{\mu}) \, ,
\end{eqnarray}
where ${\rm d}\tau=\sqrt{1-v^2/c^2}{\rm d}t$ is the proper time
element, $[\tau_i,\tau_f]$ is the proper time range and
$A^{\mu}=(\phi/c,\mathbf{A}$). The corresponding equations of
motion are:
\begin{eqnarray}\label{RelEOM2}
\frac{\rm d}{{\rm
d}t}\left[\frac{m_0\mathbf{v}}{\sqrt{1-\frac{v^2}{c^2}}}\right]=
q\left[-\nabla\phi-\frac{\partial}{\partial t}\mathbf{A} +
\mathbf{v}\times(\nabla \times \mathbf{A})\right].
\end{eqnarray}
A comparison of the right-hand side of this equation with the Lorentz
force leads to the identifications
$\mathbf{E}=-\nabla\phi-\frac{\partial}{\partial t}\mathbf{A}$ and
$\mathbf{B}=\nabla\times\mathbf{A}$, from which the homogeneous
Maxwell equations follow immediately, by using standard vector
calculus identities.

Remarkably, in \cite{FeynmanHughes2} (p. 42) there is also a hint
to a possible extension of this derivation to gravity, by noticing
that a term containing what Feynman calls a ``10-potential''
(which is a symmetric tensor) could be added as well. In such a
case, the action should read:
\begin{eqnarray}\label{action} S=-m_0c^2\int_{\tau_i}^{\tau_f} \left( \drm\tau
+ \frac{1}{2c^2} h_{\mu\nu}\frac{\drm x^{\mu}}{\drm\tau}
\frac{\drm x^{\nu}}{\drm\tau}\drm\tau\right) ,
\end{eqnarray}
and Feynman commented: ``An example of a force derivable from that
action is gravity." Clearly, the 10-potential $h$ is the
gravitational potential, which in general relativity is given by
the deviation of the metric from flat space, i.e.
$g_{\mu\nu}=\eta_{\mu\nu}+h_{\mu\nu}$, where $\eta_{\mu\nu}$ is
the Minkowski metric; however, Feynman wrote generically $g$ in
place of $h$ in the above action. In fact, we found that the
correct interaction of the particle with the gravitational field
in the weak field approximation ($|h_{\mu\nu}|\ll 1$) is given by
the action in Eq. (\ref{action}), since it can be easily obtained
from that for a particle in a curved spacetime in that
approximation:
\begin{eqnarray} S_{\text{curved}}&=&-m_0c\int_{\tau_i}^{\tau_f} ds=-m_0c\int
\sqrt{g_{\mu\nu}dx^{\mu}dx^{\nu}}= -m_0c\int_{\tau_i}^{\tau_f}
\sqrt{(\eta_{\mu\nu}+h_{\mu\nu})dx^{\mu}dx^{\nu}}\nonumber\\
&=&-m_0c\int_{\tau_i}^{\tau_f}
\sqrt{\left(\eta_{\mu\nu}+h_{\mu\nu}\right)\frac{dx^{\mu}}{d\tau}\frac{dx^{\nu}}{d\tau}}d\tau=
-m_0c^2\int_{\tau_i}^{\tau_f}
\sqrt{1+\frac{1}{c^2}h_{\mu\nu}\frac{dx^{\mu}}{d\tau}\frac{dx^{\nu}}{d\tau}}d\tau\nonumber
\\ &\approx& -m_0c^2\int_{\tau_i}^{\tau_f} \left( d\tau + \frac{1}{2c^2}
h_{\mu\nu}\frac{dx^{\mu}}{d\tau}
\frac{dx^{\nu}}{d\tau}d\tau\right)=S,
\end{eqnarray}
where in the fifth equality the fact that
$\eta_{\mu\nu}\frac{dx^{\mu}}{d\tau}\frac{dx^{\nu}}{d\tau}=c^2$
has been used. Furthermore the factor ${1}/{2}$ has been included
in (\ref{action}) in order to allow the identification of the
10-potential components $h_{\mu\nu}$ with the usual perturbations
of the metric. We also notice that the $m_0$ factor appearing in
front of the action is the inertial mass, which coincides with the
gravitational mass appearing in (\ref{Force}). Feynman, however,
did not develop this approach, likely because it would have been
very cumbersome.

\section{Conclusions}

In this paper we have reviewed Feynman's work on gravity by
considering all known published sources, as well as a few
unpublished ones, including the Hughes lectures
\cite{FeynmanHughes2}, which were made available only recently. An
emerging feature is his belief in the innermost unity of nature,
which, at its deepest level, has to be quantum. This view is
reflected in the statement that the fundamental interactions that
we experience at the macroscopic level are manifestations of
underlying quantum theories, and general relativity can indeed be
obtained (as the Maxwell theory) from the fundamental principles
of Lorentzian quantum field theory. In this approach, quantum
corrections are included as loop diagrams, with several emerging
difficulties, which Feynman tackled and partly solved. He usually
started from the field theoretical viewpoint to make even
classical computations in gravity, such as the orbits of planets
in Ref. \cite{Feynman:1996kb} and the radiation of gravitational
waves in Ref. \cite{WeisskopfLetter}, but the Hughes lectures
\cite{FeynmanHughes2} (and also the final chapter of volume II of
Ref. \cite{Feynman:1963uxa}) prove that he could also put aside
that approach in favor of a more conventional one, if the audience
required it. Even then, however, his originality emerges clearly.
On the flip side, his strategies were not always the simplest
possible, and sometimes tended to be quite cumbersome. An example
is his derivation of the equations of general relativity, which
was quite long, complicated, and not completely general, and was
later surpassed by Deser's more elegant one. And along with
Deser's and other analogous approaches, Feynman's one has some
shortcomings such as hidden assumptions (as discussed e.g. in
\cite{Padmanabhan:2004xk}). Also his route to the quantization of
gravity and Yang-Mills theory was much more difficult than the
Faddeev-Popov one, even if some tools he developed, such as the
tree theorem, have been recently found to be of wide use. Finally,
the approach to the gravitational interaction presented in Section
7.1, while being very original as its electromagnetic counterpart,
is not very useful in practice, hence Feynman did not pursue it.
Despite these drawbacks, it can be certainly said that Feynman's
work on gravity constitutes a further demonstration of his
versatility and ability to contribute substantially to almost
every branch of theoretical physics.

\section*{Acknowledgments}

The authors would like to thank the staff of the Caltech archives
for providing them a copy of Feynman's letter to Weisskopf
\cite{WeisskopfLetter}, and the anonymous referees, which provided
very useful advice on how to improve the present paper.


\begin{thebibliography}{99}

\bibitem{Feynman:1963uxa}
  R.P.~Feynman, R.B.~Leighton and M.~Sands,
  {\it The Feynman Lectures on Physics}, Addison-Wesley, Reading, MA,
  1963.

\bibitem{ChapelHill}
C. DeWitt-Morette and D. Rickles, {\it The Role of Gravitation in
Physics}, Report from the 1957 Chapel Hill Conference, Edition
Open Access, Berlin, 2011.

\bibitem{Feynman:1963ax}
  R.~P.~Feynman,
  \textit{Quantum theory of gravitation},
  Acta Phys.\ Polon.\  {\bf 24} (1963), 697-722.

\bibitem{Will}
C. Will,    {\it Was Einstein Right? Putting General Relativity to
the Test}, Oxford University Press, Oxford, 1986.

\bibitem{Will:1989rr}
C.~Will, \emph{The renaissance of general relativity}, in P.
Davies (ed.), \emph{The New Physics}, Cambridge University Press,
Cambridge, 1989.

\bibitem{Eisenstaedt1}
J. Eisenstaedt, \textit{La relativit\'e g\'en\'erale \`a
l'\'etiage: 1925-1955}, Arch. Hist. Exact Sci. \textbf{35} (1986),
115-185.

\bibitem{Eisenstaedt2}
J. Eisenstaedt, \textit{Trajectoires et impasses de la solution de
Schwarzschild}, Arch. Hist. Exact Sci. \textbf{37} (1987), 275-357.

\bibitem{Eisenstaedt3}
J. Eisenstaedt, \emph{The low water mark of general relativity,
1925-1955}, in D. Howard, J. Stachel (eds.), \emph{Einstein and
the History of General Relativity}, Birkh\"{a}user, Basel, 1989),
pp. 277-292.

\bibitem{BlumLalliRenn}
A.S. Blum, R. Lalli and J. Renn, {\it The Renaissance of General
Relativity: How and Why it Happened}, Annalen der Physik
\textbf{528} (2016), 344-349.

\bibitem{Gell-Mann}
M. Gell-Mann, \emph{Dick Feynman-The Guy in the Office Down the
Hall}, Phys. Today \textbf{42} (1989), 50-54.

\bibitem{Letter}
Letter from Bryce DeWitt to Agnew Bahnson, dated 15 November 1955, as
cited in \cite{ChapelHill}, p. 25.

\bibitem{Feynman:1996kb}
  R.~P.~Feynman, F.~B.~Morinigo, W.~G.~Wagner and B.~Hatfield,
  {\it Feynman Lectures on Gravitation},
  Addison-Wesley, Reading, MA, 1995.

\bibitem{Chandrasekhar:1964zz}
S.~Chandrasekhar, \textit{The Dynamical Instability of Gaseous
Masses Approaching the Schwarzschild Limit in General Relativity},
Astrophys. J. \textbf{140} (1964), 417-433. 

\bibitem{BergmannRMP}
P.~G.~Bergmann, \emph{Summary of the Chapel Hill conference}, Rev.
Mod. Phys.\ {\bf 29} (1957), 352-354.

\bibitem{FeynmanHughes1}
R.P. Feynman, {\it Astronomy, Astrophysics, and Cosmology},
lectures at the Hughes Aircraft Company, Notes taken and
transcribed by John T. Neer, 1966-67.

\bibitem{FeynmanHughes2} R.P. Feynman, {\it Electrostatics,
Electrodynamics, Matter-Waves Interacting, Relativity}, lectures
at the Hughes Aircraft Company, Notes taken and transcribed by
John T. Neer, 1967-68.

\bibitem{Feynmanbio}
M. Di Mauro, S. Esposito and A. Naddeo, \emph{When Physics Meets
Biology: A Less Known Feynman}, Transversal Int. J. Hist. Sci.
\textbf{4} (2018) 163-173.

\bibitem{Hughes}
\url{http://www.thehugheslectures.info/the-lectures/}, accessed on
1 August 2021.

\bibitem{DeLuca:2019ija}
  R.~De Luca, M.~Di Mauro, S.~Esposito and A.~Naddeo,
  \textit{Feynman's different approach to electromagnetism},
  Eur.\ J.\ Phys.\  {\bf 40} (2019) 065205.

\bibitem{DiMauro:2020bpd}
M.~Di Mauro, R.~De Luca, S.~Esposito and A.~Naddeo, \emph{Some
insight into Feynman's approach to electromagnetism}, Eur. J.
Phys. {\bf 42} (2021) 025206.

\bibitem{Bern}
A. Mercier and M. Kervaire (eds.), {\it F\"{u}nfzig Jahre
Relativit\"{a}tstheorie}, Birkh\"{a}user, Basel, 1956.

\bibitem{Everett}
H. Everett III, \emph{``Relative state'' formulation of quantum
mechanics}, Rev. Mod. Phys. {\bf 29} (1957) 454-462.

\bibitem{DeWittMorette:2011zz}
C.~DeWitt-Morette, \emph{The pursuit of quantum gravity: Memoirs
of Bryce DeWitt from 1946 to 2004}, Springer, Berlin, 2011.

\bibitem{DeWittBattelle}
B. S. DeWitt, \emph{The Everett-Wheeler Interpretation of Quantum
Mechanics}, in C. M. DeWitt and J. A. Wheeler (eds.)
\emph{Battelle Rencontres: 1967 Lectures in Mathematics and
Physics}, W. A. Benjamin, New York, 1968, pp 318-332.

\bibitem{DeWittPT1}
B. S. DeWitt, \emph{Quantum Mechanics and Reality}, Phys. Today
23(9) (1970), 30-35.

\bibitem{DeWittPT2}
B. S. DeWitt, L. Ballentine, P. Pearle, E. H. Walker, M. Sachs, T.
Koga, J. Gerver, \emph{Quantum Mechanics Debate}, Phys. Today
24(4) (1971), 36-44.

\bibitem{DeWittVarenna}
B. S. DeWitt, \emph{The Many-Universes Interpretation of Quantum
Mechanics}, in \emph{Proceedings of the International School of
Physics ``Enrico Fermi'' Course IL: Foundations of Quantum
Mechanics}, Academic Press, New York (1971), pp 211-262.

\bibitem{DeWittGraham}
B. S. DeWitt, N. Graham (eds.) \emph{The Many-Worlds
Interpretation of Quantum Mechanics}, Princeton University Press,
Princeton, 1973.

\bibitem{Kaiser}
D. Kaiser, \emph{A $\psi$ is just a $\psi$? Pedagogy, Practice and
the Reconstitution of General Relativity, 1942-1975}, Stud. Hist.
Phil. Mod. Phys. \textbf{29} (1998), 321-338.

\bibitem{Kennefick:2007zz}
D.~Kennefick, \emph{Traveling at the speed of thought:
Einstein and the quest for gravitational waves}, Princeton
University Press, Princeton, 2007.

\bibitem{Einstein:1916cc}
A.~Einstein, {\it  N\"{a}herungsweise Integration der
Feldgleichungen der Gravitation (Approximate Integration of the
Field Equations of Gravitation)}, Sitzungsber. Preuss. Akad. Wiss.
Berlin (Math. Phys.) \textbf{1916} (1916), 688-696.

\bibitem{Einstein:1918btx}
A.~Einstein, \textit{\"Uber Gravitationswellen (On Gravitational
Waves)}, Sitzungsber. Preuss. Akad. Wiss. Berlin (Math. Phys.)
\textbf{1918} (1918), 154-167.


\bibitem{Eddington}
A. S. Eddington, {\it The Mathematical Theory of Relativity},
2nd ed, Cambridge University Press, Cambridge, 1960.

\bibitem{Eddington:1922ds}
A.~S.~Eddington, \emph{The propagation of gravitational waves},
Proc. Roy. Soc. Lond. A \textbf{102} (1922), 268-282.

\bibitem{Pais}
A. Pais, {\it Subtle is the Lord. The Science and Life of
Albert Einstein}, Oxford University Press, Oxford, 1982.

\bibitem{EinsteinBorn}
A. Einstein and M. Born, {\it The Born-Einstein Letters:
Friendship, Politics, and Physics in Uncertain Times}, MacMillan,
New York, 2005.

\bibitem{EinsteinRosen1937}
A. Einstein and N. Rosen, \emph{On Gravitational waves}, J. Frank.
Inst.\ {\bf 223} (1937), 43-54.

\bibitem{Pirani1}
F. A. E. Pirani, \emph{On the Physical significance of the Riemann
tensor}, Acta Phys. Pol.\ {\bf 15} (1956), 389-405; reprinted in Gen.
Rel. Grav.\ {\bf 41} (2009), 1215-1232.

\bibitem{Pirani2}
F. A. E. Pirani, \emph{Invariant formulation of gravitational
radiation theory}, Phys. Rev. {\bf 105} (1957), 1089-1098.

\bibitem{Weber:1957oib}
J.~Weber and J.~A.~Wheeler, \emph{Reality of the Cylindrical
Gravitational Waves of Einstein and Rosen},
Rev. Mod. Phys. \textbf{29} (1957), 509-515.

\bibitem{postN1}
L. Blanchet, \emph{Gravitational Radiation from Post-Newtonian
Sources and Inspiralling Compact Binaries}, Living Rev. Relativity
\ {\bf 9} (2006) 4.

\bibitem{postN2}
T. Futamase and Y. Itoh, \emph{The Post-Newtonian Approximation for Relativistic Compact Binaries},
 Living Rev. Relativity \ {\bf 10} (2007) 2.

\bibitem{Landau}
L. D. Landau and E. M. Lifshitz, \emph{The Classical Theory
of Fields,} first English edition, Addison- Wesley, Cambridge,
1951.

\bibitem{FeynmanWheeler1}
R. P. Feynman, J. A. Wheeler, \emph{Reaction of the Absorber as
the Mechanism of Radiative Damping} (Abstract only), Phys. Rev.
\textbf{59} (1941), 692.

\bibitem{FeynmanWheeler2}
J. A. Wheeler, R. P. Feynman, \emph{Interaction with the absorber
as the mechanism of radiation }, Rev. Mod. Phys.
\textbf{17}(1945), 157-181.

\bibitem{FeynmanWheeler3}
J. A. Wheeler, R. P. Feynman, \emph{Classical electrodynamics in
terms of direct interparticle action }, Rev. Mod. Phys.
\textbf{21} (1945), 425-433.

\bibitem{WeisskopfLetter}
R. P. Feynman, unpublished letter to Victor F. Weisskopf, February 1961; in Richard P. Feynman Papers, California Institute of Technology Archives, Box 3, Folder 8.

\bibitem{Bondi}
H. Bondi, \emph{Plane gravitational waves in general relativity},
Nature\ {\bf 179} (1957), 1072-1073.

\bibitem{Bondi:1958aj}
H.~Bondi, F.~A.~E.~Pirani and I.~Robinson, \emph{Gravitational
waves in general relativity. 3. Exact plane waves}, Proc. Roy.
Soc. Lond. A \textbf{251} (1959), 519-533.

\bibitem{Robinson:1960zzb}
I.~Robinson and A.~Trautman, \emph{Spherical Gravitational Waves},
Phys. Rev. Lett. \textbf{4} (1960), 431-432.

\bibitem{Bondi:1962px}
H.~Bondi, M.~G.~J.~van der Burg and A.~W.~K.~Metzner,
\emph{Gravitational waves in general relativity. 7. Waves from
axisymmetric isolated systems}, Proc. Roy. Soc. Lond. A
\textbf{269} (1962), 21-52.

\bibitem{Sachs:1962wk}
R.~K.~Sachs, \emph{Gravitational waves in general relativity. 8.
Waves in asymptotically flat space-times},
Proc. Roy. Soc. Lond. A \textbf{270} (1962), 103-126.

\bibitem{Sachs:1964zza}
R.~K.~Sachs, \emph{Gravitational radiation}, In \emph{Relativity,
Groups and Topology}, C. DeWitt and B. S. DeWitt eds., Gordon and
Breach, New York, 1964, pp. 521-562.

\bibitem{Newman:1961qr}
E.~Newman and R.~Penrose, \emph{An Approach to gravitational
radiation by a method of spin coefficients},
J. Math. Phys. \textbf{3} (1962), 566-578. 

\bibitem{Penrose:1965am}
R.~Penrose, \emph{Zero rest mass fields including gravitation:
Asymptotic behavior}, Proc. Roy. Soc. Lond. A \textbf{284} (1965), 159-203.

\bibitem{Weber1}
J. Weber, \emph{Detection and Generation of Gravitational Waves},
Phys. Rev.\ {\bf 117} (1960), 306-313.

\bibitem{Weber2}
J. Weber, \emph{Observation of the Thermal Fluctuations of a
Gravitational-Wave Detector}, Phys. Rev. Lett.\ {\bf 17} (1966), 1228-1230.

\bibitem{Weber3}
J. Weber, \emph{ Evidence for discovery of gravitational
radiation}, Phys. Rev. Lett.\ {\bf 22} (1969), 1320-1324.

\bibitem{Zeh}
H.~D.~Zeh, \emph{Feynman's interpretation of quantum theory}, Eur.
Phys. J. H\  {\bf 36} (2011), 63-74.

\bibitem{VonN}
J. von Neumann, {\it Mathematische Grundlagen der Quantenmechanik
(Mathematical foundations of quantum mechanics)}, Springer,
Berlin, 1932 (Chap. 6).

\bibitem{ZurekRev}
W. H. Zurek, \emph{Decoherence, einselection, and the quantum
origins of the classical}, Rev. Mod. Phys.\ {\bf 75} (2003), 715-776.

\bibitem{karol}
F. K\'arolyh\'azy, \emph{Gravitation and quantum mechanics of
macroscopic objects}, Il Nuovo Cimento A {\bf 42} (1966), 390-402.

\bibitem{frenkel}
A. Frenkel, \emph{Spontaneous localizations of the wave function
and classical behavior}, Found. Phys. {\bf 20} (1990), 159-188.


\bibitem{diosi0}
L. Diosi, \emph{Gravitation and quantum mechanical localization of
macroobjects}, Phys. Lett. A {\bf 105} (1984), 199-202.

\bibitem{diosi}
L. Diosi, \emph{Models for universal reduction of macroscopic
quantum fluctuations}, Phys. Rev. A {\bf 40} (1989), 1165-1174.

\bibitem{diosi1}
L. Diosi, \emph{Notes on certain Newton gravity mechanisms of
wavefunction localization and decoherence}, J. Phys. A: Math. Gen.
{\bf 40} (2007), 2989-2995.

\bibitem{penrose0}
R. Penrose, \emph{Gravity and state-vector reduction}, in R.
Penrose and C. J. Isham (Eds.), \textit{Quantum Concepts in Space
and Time}, Clarendon Press, Oxford, 1986 (p. 129).

\bibitem{penrose1}
R. Penrose, \emph{On gravity's role in quantum state reduction}
Gen. Rel. Grav. {\bf 28} (1996), 581-600.

\bibitem{penrose2}
R. Penrose, \emph{Quantum computation, entanglement and state
reduction}, Phil. Trans. R. Soc. Lond. A {\bf 356} (1998), 1927-1939.

\bibitem{penrose3}
R. Penrose, \textit{The Road to Reality: A Complete Guide to the
Laws of the Universe}, Jonathan Cape, London, 2004.

\bibitem{random1}
G. C. Ghirardi, R. Grassi and A. Rimini, \emph{Continuous
spontaneous reduction model involving gravity}, Phys. Rev. A {\bf 42} (1990), 1057-1064.

\bibitem{random2}
G. C. Ghirardi, A. Rimini and T. Weber, \emph{Unified Dynamics for Microscopic and Macroscopic Systems}, Phys. Rev. D {\bf 34} (1986), 470-491.

\bibitem{random3}
P. Pearle, \emph{Ways to describe dynamical state-vector
reduction}, Phys. Rev. A {\bf 48} (1993), 913-923.

\bibitem{random4}
A. Bassi and G. C. Ghirardi, \emph{Dynamical reduction models},
Phys. Rept. {\bf 379} (2003), 257-426.

\bibitem{arndt1}
S. Gerlich, S. Eibenberger, M. Tomandl, S. Nimmrichter, K.
Hornberger, P. J. Fagan, J. Tuxen, M. Mayor and M. Arndt,
\emph{Quantum interference of large organic molecules}, Nat.
Commun. {\bf 2} (2011), 263.

\bibitem{arndt2}
S. Eibenberger, S. Gerlich, M. Arndt, M. Mayor and J. Tuxen,
\emph{Matter-wave interference of particles selected from a
molecular library with masses exceeding 10 000 amu}, Phys. Chem.
Chem. Phys. {\bf 15} (2013), 14696-14700.

\bibitem{matterW1}
I. Pikovski, M. Zych, F. Costa and C. Brukner, \emph{Universal
decoherence due to gravitational time dilation}, Nat. Phys. {\bf
11} (2015), 668-672.

\bibitem{matterW2}
M. Zych, F. Costa, I. Pikovski and C. Brukner, \emph{Quantum
interferometric visibility as a witness of general relativistic
proper time}, Nat. Commun. {\bf 2} (2011), 505.

\bibitem{matterW3}
Y. Margalit, Z. Zhou, S. Machluf, D. Rohrlich, Y. Japha and R.
Folman, \emph{A self-interfering clock as a ``which path''
witness}, Science {\bf 349} (2015), 1205-1208.

\bibitem{bose}
S. Bose, K. Jacobs and P. L. Knight, \emph{Scheme to probe the
decoherence of a macroscopic object}, Phys. Rev. A {\bf 59} (1999), 3204-3210.

\bibitem{marshall}
W. Marshall, C.  Simon, R. Penrose and D. Bouwmeester,
\emph{Towards quantum superpositions of a mirror} Phys. Rev.
Lett. {\bf 91} (2003), 130401.

\bibitem{mio}
F. Maimone, G. Scelza, A. Naddeo and V. Pelino, \emph{Quantum
superpositions of a mirror for experimental tests for nonunitary
Newtonian gravity}, Phys. Rev. A {\bf 83} (2011), 062124.

\bibitem{oriol}
O. Romero-Isart, A. C. Pflanzer, F. Blaser, R. Kaltenbaek, N.
Kiesel, M. Aspelmeyer and J. I. Cirac, \emph{Large Quantum
Superpositions and Interference of Massive Nanometer-Sized
Objects}, Phys.  Rev.  Lett. {\bf 107} (2011), 020405.

\bibitem{mag1}
H. Pino, J. Prat-Camps, K. Sinha, B. P. Venkatesh and O.
Romero-Isart, \emph{On-chip quantum interference of a
superconducting microsphere}, Quantum Sci. Technol. {\bf 3} (2018), 25001.

\bibitem{Osnaghi}
S. Osnaghi, F. Freitas, O. Freire Jr., \emph{The origin of the
Everettian heresy}, Stud. Hist. Phil. Mod. Phys. {\bf 40} (2009), 97-123.

\bibitem{PreskillTalk}
J. Preskill, \emph{Feynman after 40}, Talk given at APS April
Meeting, 16 April 2018, available at
\url{http://theory.caltech.edu/~preskill/talks/APS-April-2018-Feynman-4-3.pdf},
accessed on 22 May 2021.

\bibitem{dh1}
M. Gell-Mann and J. B. Hartle, \emph{Classical equations for quantum
systems}, Phys. Rev. D {\bf 47} (1993), 3345-3382.

\bibitem{dh2}
R. B. Griffiths, \textit{Consistent Quantum Theory}, Cambridge
University Press, Cambridge, 2002.

\bibitem{dh3}
R. Omnes, \textit{Interpretation of Quantum Mechanics}, Princeton
University Press, Princeton, 1994.

\bibitem{dh4}
M. Gell-Mann and J. B. Hartle, \emph{Quantum Mechanics in the
Light of Quantum Cosmology}, in W. H. Zurek (Ed.),
\textit{Complexity, Entropy, and the Physics of Information}, SFI
Studies in the Sciences of Complexity, Vol. VIII, Addison Wesley,
Reading, MA, 1990.

\bibitem{dh5}
P. C. Hohenberg, \emph{An Introduction to consistent quantum
theory}, Rev. Mod. Phys. {\bf 82} (2010), 2835-2844.

\bibitem{dh6}
J. B. Hartle, \emph{The Quantum Mechanics of Closed Systems}, in
B. L. Hu, M. P. Ryan, C. V. Vishveshwara (Eds.),
\textit{Directions in General Relativity}, Volume 1, Cambridge
University Press, Cambridge, 1993.

\bibitem{dh7}
J. B. Hartle, \emph{The Impact of Cosmology on Quantum Mechanics},
[arXiv:1901.03933 [gr-qc]].


\bibitem{Kraichnan:1955zz}
  R.~H.~Kraichnan,
  \emph{Special-Relativistic Derivation of Generally Covariant Gravitation
  Theory},
  Phys.\ Rev.\  {\bf 98} (1955), 1118-1122.

\bibitem{Gupta:1954zz}
  S.~N.~Gupta,
  \emph{Gravitation and Electromagnetism},
  Phys.\ Rev.\  {\bf 96} (1954), 1683-1685.

\bibitem{LectGravPreface}
J. Preskill and K. S. Thorne, Foreword to Ref. \cite{Feynman:1996kb}.

\bibitem{Fierz:1939ix}
  M.~Fierz and W.~Pauli,
  \emph{On relativistic wave equations for particles of arbitrary spin in an electromagnetic field},
  Proc.\ Roy.\ Soc.\ Lond.\ A {\bf 173} (1939), 211-232.

\bibitem{Dirac:1936tg}
  P.~A.~M.~Dirac,
  \emph{Relativistic wave equations},
  Proc.\ Roy.\ Soc.\ Lond.\ A {\bf 155} (1936), 447-459.

\bibitem{Padmanabhan:2004xk}
  T.~Padmanabhan,
  \emph{From gravitons to gravity: Myths and reality},
  Int.\ J.\ Mod.\ Phys.\ D {\bf 17} (2008), 367-398.

\bibitem{Rickles1}
D. Rickles, \textit{Covered with deep mist}, Oxford University
Press, Oxford, 2020.

\bibitem{RicklesBlum}
A. S. Blum and D. Rickles, {\it Quantum Gravity in the First Half
of the Twentieth Century: A Sourcebook}, Edition Open Access,
Berlin, 2018.

\bibitem{HartzBlum}
A. Blum, T. Hartz, \emph{The 1957 quantum gravity meeting in
Copenhagen: An analysis of Bryce S. DeWitt's report}, Eur. Phys.
J. H \textbf{42} (2017), 107-157.

\bibitem{HeiPau1}
W. Heisenberg and W. Pauli, \emph{Zur Quantenelektrodynamik der
Wellenfelder (On the quantum electrodynamics of wave fields)},
Zeitschrift fur Physik \textbf{56} (1929), 1-61.

\bibitem{Ros1}
L. Rosenfeld, \emph{Zur Quantelung der Wellenfelder (On the
quantization of wave fields}, Annalen der Physik \textbf{5} (1930)
1113-1152. English translation by D. C. Salisbury and K.
Sundermeyer, Eur. Phys. J H \textbf{42} (2017), 63-94.

\bibitem{Ros2}
L. Rosenfeld, \emph{\"{U}ber die Gravitationswirkungen des Lichtes
(On the gravitational effects of light)}, Zeitschrift fur Physik
\textbf{65} (1930), 589-599.


\bibitem{Peruzzi:2018nzo}
G.~Peruzzi and A.~Rocci, \emph{Tales from the prehistory of
Quantum Gravity: L\'eon Rosenfeld\textquoteright{}s earliest
contributions}, Eur. Phys. J. H \textbf{43} (2018), 185-241.

\bibitem{Salisbury1}
D. C. Salisbury, \emph{Leon Rosenfeld's pioneering steps toward a
quantum theory of gravity}, J. Phys. Conf. Ser. \textbf{222}
(2010), 012052.

\bibitem{Salisbury2}
D. C. Salisbury, K. Sundermeyer, \emph{Leon Rosenfeld's general
theory of constrained Hamiltonian dynamics},  Eur. J. Phys. H
\textbf{42} (2017), 23-61.

\bibitem{Salisbury:2009cr}
D.~Salisbury, \emph{Leon Rosenfeld and the challenge of the
vanishing momentum in quantum electrodynamics}, Stud. Hist. Phil.
Sci. B \textbf{40} (2009), 363-373.

\bibitem{Stachel}
J. Stachel, \emph{The early history of quantum gravity
(1916-1940)'}, in B. R. Iyer and B. Bhawal (Eds.), \textit{Black
Holes, Gravitational radiation and the Universe}, Kluwer, The
Netherlands, 1999.

\bibitem{Galperin}
D. I. Blokhintsev and F. M. Gal'perin, \emph{Gipoteza neitrino i
zakon sokhraneniya energii (Neutrino hypothesis and energy
conservation)}, Pod znamenem marxisma \textbf{6} (1934), 147-157.

\bibitem{Bronstein}
M. P. Bronstein, \emph{Quantentheories schwacher
Gravitationsfeldern (Quantum theory of weak gravitational
fields)}, Physikalische Zeitschrift der Sowietunion \textbf{9}
(1936), 140-157. English translation by M. A. Kurkov, edited by S.
Deser, Gen. Relativ. Grav. \textbf{44} (2012), 267-283.

\bibitem{Gorelik:2005an}
G.~E.~Gorelik, \emph{Matvei Bronstein and quantum gravity: 70th
anniversary of the unsolved problem}, Phys. Usp. \textbf{48}
(2005), 1039-1053.

\bibitem{GorelikBook}
G. E. Gorelik, V. Ya. Frenkel, \emph{Matvei Petrovich Bronstein
and Soviet theoretical physics in the sixties,} Birkh\"{a}user,
1994.

\bibitem{BronsteinComment}
S. Deser, A. Starobinsky, \emph{Editorial note to: Matvei P.
Bronstein, Quantum theory of weak gravitational fields}, Gen.
Relativ. Grav. \textbf{44} (2012), 263-265.

\bibitem{Fermi:1932xva}
E.~Fermi, \emph{Quantum theory of radiation}, Rev. Mod.
Phys.\textbf{4} (1932), 87-132.

\bibitem{BohrRosenfeld}
N. Bohr, L. Rosenfeld, \emph{Zur Frage der Messbarkeit der
elektromagnetischen Feldgr\"{o}\ss en (On the question of
measurability of electromagnetic field quantities)}, Det Kgl.
Danske Videnskabernes Selskab Mathematisk-fysiske Meddelelser
\textbf{12} (1933), 3-65.

\bibitem{Solomon}
J. Solomon, \emph{Gravitation et Quanta}, Journal de Physique et
de Radium \textbf{9} (1938), 479-485.

\bibitem{Rosen}
N. Rosen\, \emph{Plane Polarised Waves in the General Theory of
Relativity}, Phys. Zeitsch. der Sowjetunion \textbf{12}, 366-372.

\bibitem{DeWittPHD}
B. DeWitt, \emph{I: The Theory of Gravitational Interactions. II:
The Interaction of Gravitation with Light}, Ph.D. thesis, Harvard,
1949.

\bibitem{Schwinger}
J. Schwinger, \emph{Quantum electrodynamics. I. A covariant
formulation}, Phys. Rev. \textbf{74} (1948), 1439-1461.

\bibitem{Berg1}
P. G. Bergmann, \emph{Non-linear Field Theories}, Phys. Rev.
\textbf{75} (1949), 680-685.

\bibitem{Berg2}
P. G. Bergmann, J. H. M. Brunings, \emph{Non-linear Field Theories
II: Canonical Equations and Quantization}, Rev. Mod. Phys.
\textbf{21} (1949), 480-487.

\bibitem{Salisbury:2012ona}
D.~C.~Salisbury, \emph{Peter Bergmann and the invention of
constrained Hamiltonian dynamics}, Einstein Stud. \textbf{12}
(2012), 247-257.

\bibitem{Salisbury:2007br}
D.~C.~Salisbury, \emph{Rosenfeld, Bergmann, Dirac and the
invention of constrained Hamiltonian dynamics}, in H. Kleinert, R.
T. Jantzen, R. Ruffini (eds.), \emph{Proceedings of the eleventh
Marcel Grossmann meeting on general relativity}, World Scientific,
Singapore, 2008.

\bibitem{DiracHam}
P. A. M. Dirac, \emph{Generalized Hamiltonian dynamics}, Canadian
J. Math. \textbf{2} (1950), 129-148.

\bibitem{DiracForms}
P.~A.~M.~Dirac, \emph{Forms of Relativistic Dynamics}, Rev. Mod.
Phys. \textbf{21} (1949), 392-399.

\bibitem{Schild}
F. A. E. Pirani and A. Schild, \emph{On the quantization of
Einstein's gravitational field equations}, Phys. Rev. \textbf{79}
(1950), 986-991.

\bibitem{WheelerFest2}
R. P. Feynman, \emph{Problems in quantizing the gravitational
field and the massless Yang-Mills field}, in Ref.
\cite{Klauder:1972je}.

\bibitem{Weinberg:1964kqu}
S.~Weinberg, \emph{Derivation of gauge invariance and the
equivalence principle from Lorentz invariance of the S- matrix},
Phys. Lett. \textbf{9} (1964), 357-359.

\bibitem{Weinberg:1965rz}
  S.~Weinberg,
  \emph{Photons and gravitons in perturbation theory: Derivation of Maxwell's and Einstein's equations},
  Phys.\ Rev.\  {\bf 138} (1965), B988-B1002.

\bibitem{Deser:1969wk}
  S.~Deser,
  \emph{Self-interaction and gauge invariance},
  Gen.\ Rel.\ Grav.\  {\bf 1} (1970), 9-18.

\bibitem{Boulware:1974sr}
  D.~G.~Boulware and S.~Deser,
  \emph{Classical General Relativity Derived from Quantum Gravity},
  Annals Phys.\  {\bf 89} (1975), 193-240.

\bibitem{Deser87}
S. Deser, \emph{Gravity From Self-interaction in a Curved
Background}, Class. Quant. Grav. \textbf{4} (1987), L99-L106.

\bibitem{Deser:2009fq}
S.~Deser, \emph{Gravity from self-interaction redux},
Gen. Rel. Grav. \textbf{42} (2010), 641-646.

\bibitem{Wald:1986bj}
R.~M.~Wald, \emph{Spin-2 Fields and General Covariance},
Phys. Rev. D \textbf{33} (1986), 3613-3625

\bibitem{Thorne:1994xa}
K.~S.~Thorne, \textit{Black holes and time warps: Einstein's
outrageous legacy}, Norton, New York, 1994.

\bibitem{Chew} G. F. Chew, \textit{S-matrix theory of strong interactions: a lecture note and reprint volume}, Frontiers in
physics series, W.A. Benjamin, New York, 1961.

\bibitem{LaJolla}
W. R. Frazer, \emph{Theory of Weak and Strong Interactions},
Physics Today {\bf 14}(12) (1961), 80-84.

\bibitem{WheelerFest1}
R. P. Feynman, \emph{Closed loop and tree diagram}, in Ref.
\cite{Klauder:1972je}.

\bibitem{Klauder:1972je}
  J.~R.~Klauder (ed.),
  \textit{Magic Without Magic - John Archibald Wheeler. A Collection Of Essays In Honor Of His 60th
  Birthday,}
  Freeman, San Francisco, 1972.

\bibitem{Einstein1933}
A. Einstein, \emph{Notes on the Origin of the General Theory of
Relativity}, In \emph{Ideas and Opinions}, Translated by Sonja
Bargmann, Crown, New York, 1954, pp. 285-290.

\bibitem{Nordstrom}
G. Nordstr\"{o}m, \emph{Relativit\"{a}tsprinzip und Gravo+itation
(The principle of relativity and gravitation)}, Phys. Z.
\textbf{13} (1912), 1126-1129.

\bibitem{Norton}
J. D. Norton, \emph{Einstein, Nordstr\"{o}m and the early demise
of scalar, Lorentz-covariant theories of gravitation}, Arch. Hist.
Exact Sci. \textbf{45} (1992), 17-94. Reprinted in J. Renn (ed.),
\emph{The Genesis of General Relativity, Vol. 3: Theories of
Gravitation in the Twilight of Classical Physics. Part I.},
Kluwer, 2005, pp. 413-487.

\bibitem{Notes}
R. P. Feynman, Physics 2C: Miscellaneous notes, Folder 40.5,
available online at
\url{https://www.feynmanlectures.caltech.edu/Notes.html}

\bibitem{Trautman}
A. Trautman, D. Salisbury, \emph{Memories of my early career in
relativity physics}, Eur. Phys. J. H \textbf{44} (2019), 391-413.

\bibitem{Britto:2010xq}
R.~Britto, \emph{Loop Amplitudes in Gauge Theories: Modern
Analytic Approaches}, J. Phys. A \textbf{44} (2011), 454006.

\bibitem{Bierenbaum:2010xg}
I.~Bierenbaum, S.~Catani, P.~Draggiotis and G.~Rodrigo,
\emph{Feynman's Tree Theorem and Loop-Tree Dualities},
PoS \textbf{LC2010} (2010), 034.

\bibitem{CaronHuot:2010zt}
S.~Caron-Huot, \emph{Loops and trees},
JHEP \textbf{05} (2011), 080. 

\bibitem{Maniatis:2016nmc}
M.~Maniatis, \emph{Application of the Feynman-tree theorem
together with BCFW recursion relations}, Int. J. Mod. Phys. A
\textbf{33} (2018) 1850042.



\bibitem{DeWitt:1964yg}
B.~S.~DeWitt, \emph{Theory of radiative corrections for
non-abelian gauge fields},
Phys. Rev. Lett. \textbf{12} (1964), 742-746.

\bibitem{DeWitt:1967ub}
B.~S.~DeWitt, \emph{Quantum Theory of Gravity. 2. The Manifestly
Covariant Theory},
Phys. Rev. \textbf{162} (1967), 1195-1239.

\bibitem{DeWitt:1967uc}
B.~S.~DeWitt, \emph{Quantum Theory of Gravity. 3. Applications of
the Covariant Theory},
Phys. Rev. \textbf{162} (1967), 1239-1256.

\bibitem{FaddeevPopov}
L. D. Faddeev and V. N. Popov, \emph{Feynman Diagrams for the
Yang-Mills Field}, Phys. Lett. B \textbf{25} (1967), 29-30.

\bibitem{Stelle:1976gc}
K.~S.~Stelle, \emph{Renormalization of Higher Derivative Quantum
Gravity}, Phys. Rev. D \textbf{16} (1977), 953-969.

\bibitem{Anselmi:2017ygm}
D.~Anselmi, \emph{On the quantum field theory of the gravitational
interactions}, JHEP \textbf{06} (2017), 086.

\bibitem{tHooft:1971akt}
G.~'t Hooft, \emph{Renormalization of Massless Yang-Mills Fields},
Nucl. Phys. B \textbf{33} (1971), 173-199.

\bibitem{tHooft:1971qjg}
G.~'t Hooft, \emph{Renormalizable Lagrangians for Massive
Yang-Mills Fields},
Nucl. Phys. B \textbf{35} (1971), 167-188.

\bibitem{tHooft:1972tcz}
G.~'t Hooft and M.~J.~G.~Veltman, \emph{Regularization and
Renormalization of Gauge Fields},
Nucl. Phys. B \textbf{44} (1972), 189-213.

\bibitem{tHooft:1972qbu}
G.~'t Hooft and M.~J.~G.~Veltman,
 \emph{Combinatorics of gauge fields},
Nucl. Phys. B \textbf{50} (1972), 318-353.

\bibitem{tHooft:1974toh}
G.~'t Hooft and M.~J.~G.~Veltman, \emph{One loop divergencies in
the theory of gravitation},
Ann. Inst. H. Poincare Phys. Theor. A \textbf{20} (1974), 69-94.

\bibitem{Goroff:1985sz}
M.~H.~Goroff and A.~Sagnotti, \emph{Quantum gravity at two loops},
Phys. Lett. B \textbf{160} (1985), 81-86.

\bibitem{Goroff:1985th}
M.~H.~Goroff and A.~Sagnotti, \emph{The Ultraviolet Behavior of
Einstein Gravity},
Nucl. Phys. B \textbf{266} (1986), 709-736.

\bibitem{vandeVen:1991gw}
A.~E.~M.~van de Ven, \emph{Two loop quantum gravity},
Nucl. Phys. B \textbf{378} (1992), 309-366.

\bibitem{Mehra:1994dz}
J.~Mehra, \emph{The Beat of a different drum: The Life and
science of Richard Feynman}, Oxford University Press, Oxford, 1994.

\bibitem{Weinberg:2021exr}
S.~Weinberg, \emph{On the Development of Effective Field Theory},
Eur. Phys. J. H \textbf{46} (2021), 6.

\bibitem{Weinberg:2016kyd}
S.~Weinberg, \emph{Effective field theory, past and future}, Int.
J. Mod. Phys. A \textbf{31} (2016), 1630007.

\bibitem{Burgess:2003jk}
C.~P.~Burgess, \emph{Quantum gravity in everyday life: General
relativity as an effective field theory},
Living Rev. Rel. \textbf{7} (2004), 5-56 



\bibitem{RohrlichBook}
F. Rohrlich, \emph{Classical Charged Particles}, 3rd ed., World
Scientific, Singapore 2007.

\bibitem{FultonRohrlich}
T. Fulton and F. Rohrlich, \emph{Classical radiation from a uniformly
accelerated charge}, Ann. Phys. \textbf{9} (1960), 499-517.

\bibitem{Rohrlich}
F. Rohrlich, \emph{The principle of equivalence}, Ann. Phys.
\textbf{22} (1963), 169-191.

\bibitem{DeWitt:1960fc}
B.~S.~DeWitt and R.~W.~Brehme, \emph{Radiation damping in a
gravitational field},
Ann. Phys. \textbf{9} (1960), 220-259.

\bibitem{DeWitt:1964de}
C.~M.~DeWitt and B.~S.~DeWitt, \emph{Falling charges}, Physics
\textbf{1} (1964), 3-20; err. ibid. \textbf{1} (1964), 145.

\bibitem{Boulware}
D. G. Boulware, \emph{Radiation from a Uniformly Accelerated
Charge}, Ann. Phys. \textbf{124} (1980), 169-188.

\bibitem{deAlmeida}
C. de Almeida, and A. Saa, \emph{The radiation of a uniformly
accelerated charge is beyond the horizon: A simple derivation},
Am. J. Phys. \textbf{74} (2006), 154-158.

\bibitem{WeberBook}
J. Weber, \emph{General Relativity and Gravitational Waves},
Interscience Publishers, New York, 1961.

\bibitem{Einstein1911}
A. Einstein, \emph{Einfluss der Schwerkraft auf die Ausbreitung
des Lichtes (On the influence of gravitation on the propagation of
light)}, Ann. Phys. \textbf{35} (1911), 898-908.

\bibitem{Einstein1916}
A. Einstein, \emph{Grundlage der allgemeinen Relativit\"atstheorie
(The foundation of the general theory of relativity)},  Ann.
Phys. \textbf{49} (1916), 769-822.

\bibitem{Gottlieb}
R.P. Feynman, \emph{Working Notes and Calculations: Alternate Way
to Handle Elctrodynamics}; December 13, 1963, Richard P. Feynman
Papers, California Institute of Technology Archives, Box 62,
Folder 8. Notes scanned and transcribed by M. A. Gottlieb,
available online at
\url{http://www.feynmanlectures.caltech.edu/info/other/Alternate_Way_to_Handle_Electrodynamics.html},
accessed on 16 May 2021.

\bibitem{Krasnov}
K. Krasnov, \emph{Formulations of general relativity. Gravity,
spinors and differential forms}, Cambridge University Press,
Cambridge, 2000.

\end{thebibliography}
\end{document}